\documentclass[11pt]{scrartcl}

\usepackage[margin=2.0cm]{geometry}

\usepackage[utf8]{inputenc}
\usepackage[T1]{fontenc}
\usepackage[english]{babel}

\usepackage{amsmath,amssymb,amsfonts}
\usepackage{amsthm}
\usepackage{mathrsfs}
\usepackage[title]{appendix}
\usepackage{bm}
\usepackage{booktabs}
\usepackage{algorithm}
\usepackage{algorithmicx}
\usepackage{algpseudocode}
\usepackage{cancel}
\usepackage[format=plain,singlelinecheck=false,font={small,bf},labelfont=bf,labelsep=period]{caption}
\DeclareCaptionFont{xbf}{\small\bfseries\boldmath}
\usepackage{graphicx}
\usepackage{float}
\usepackage{color}
\usepackage{xcolor}
\usepackage{colortbl}
\usepackage{csquotes}
\usepackage{dcolumn}
\usepackage{fancyhdr}
\usepackage{helvet}
\usepackage{listings}
\usepackage{lscape}
\usepackage{mathpazo}
\usepackage{mathtools}
\usepackage[version=3]{mhchem}
\usepackage{microtype}
\usepackage{multirow}
\usepackage{pgfplots}
\usepackage{setspace}
\usepackage{siunitx}
\usepackage{subfigure}
\usepackage{tablefootnote}
\usepackage{txfonts}
\usepackage[normalem]{ulem}
\usepackage{xkeyval}
\usepackage{textcomp}
\usepackage{manyfoot}
\DeclareUnicodeCharacter{2212}{-}

\RedeclareSectionCommand[
  beforeskip=-3.0ex plus -1ex minus -.2ex,
  afterskip=1.5ex plus .2ex
]{section}
\RedeclareSectionCommand[
  beforeskip=-2.0ex plus -0.8ex minus -.2ex,
  afterskip=1.0ex plus .2ex
]{subsection}

\usepackage{xr}
\usepackage{hyperref}
\hypersetup{
  breaklinks=true,
  colorlinks=true,
  citecolor=blue,
  linkcolor=blue,
  filecolor=blue,
  urlcolor=blue
}
\usepackage{cleveref}

\usepackage[numbers,sort&compress]{natbib}
\setcitestyle{super}
\usepackage{natmove}

\makeatletter
\@ifundefined{abx@aux@refcontext}{%
  \newcommand\abx@aux@refcontext[2]{}%
}{}
\@ifundefined{abx@aux@sortscheme}{%
  \newcommand\abx@aux@sortscheme[1]{}%
}{}
\@ifundefined{abx@aux@cite}{%
  \newcommand\abx@aux@cite[3]{}%
}{}
\@ifundefined{abx@aux@segm}{%
  \newcommand\abx@aux@segm[5]{}%
}{}
\makeatother

\usepackage{authblk}

\usepackage{letltxmacro}
\LetLtxMacro{\oldsqrt}{\sqrt}
\renewcommand{\sqrt}[2][\mkern8mu]{\mkern-8mu\mathop{}\oldsqrt[#1]{#2}}

\def\SM{Supporting Information}

\title{Teaching a Transformer to Think Like a Chemist: Predicting Nanocluster Stability}

\author[1]{Jo{\~a}o Marcos T. Palheta}
\author[1]{Octavio Rodrigues Filho}
\author[2]{Mohammad Soleymanibrojeni}
\author[3]{Alexandre Cavalheiro Dias}
\author[4]{Diego Guedes-Sobrinho}
\author[5]{Wolfgang Wenzel}
\author[2]{Roland Aydin}
\author[5]{Celso R. C. R\^ego\thanks{Corresponding author: \texttt{celso.rego@kit.edu}}}
\author[1]{Maur\'icio Jeomar Piotrowski}

\affil[1]{Department of Physics, Federal University of Pelotas, 96010-900 Pelotas, Rio Grande do Sul, Brazil}
\affil[2]{Hamburg University of Technology, 21073 Hamburg, Germany}
\affil[3]{Institute of Physics and International Center of Physics, University of Bras{\'i}lia, 70919-970 Bras{\'i}lia, Distrito Federal, Brazil}
\affil[4]{Quantum Chemistry and Thermodynamic Materials Group -- Q$^2$M, Department of Chemistry, Federal University of Paran{\'a}, 81531-980 Curitiba, Paran{\'a}, Brazil}
\affil[5]{Institute of Nanotechnology Hermann-von-Helmholtz-Platz, Karlsruhe Institute of Technology, Karlsruhe, Baden-Württemberg, Germany}

\date{} 

\begin{document}

\maketitle

\begin{abstract}
Atomically precise metal nanoclusters bridge the molecular and bulk regimes, but designing bimetallic motifs with targeted stability and reactivity remains challenging. Here we combine density functional theory (DFT) and physics-grounded predictive artificial intelligence to map the configurational landscape of 13-atom icosahedral nanoclusters \ce{X12TM}, with hosts \ce{X = (Ti, \text{ } Zr, \text{ } Hf)}, and \ce{Fe} and a single transition-metal dopant spanning the 3$d$-5$d$ series. Spin-polarized DFT calculations on \num{240} bimetallic clusters reveal systematic trends in binding and formation energies, distortion penalties, effective coordination number, d-band centre, and HOMO--LUMO gap that govern the competition between core--shell (in) and surface-segregated (out) arrangements. We then pretrain a transformer architecture on a curated set of 2968 unary clusters from the Quantum Cluster Database and fine-tune it on bimetallic data to predict formation energies and in/out preference, achieving mean absolute errors of about 0.6--0.7\SI{}{\electronvolt} and calibrated uncertainty intervals. The resulting model rapidly adapts to an unseen Fe-host domain with only a handful of labelled examples. At the same time, attention patterns and Shapley attributions highlight size mismatch, $d$-electron count, and coordination environment as key descriptors. All data, code, and workflows follow FAIR/TRUE principles, enabling reproducible, interpretable screening of unexplored nanocluster chemistries for catalysis and energy conversion.
\end{abstract}


\section{Introduction}\label{sec1}

Atomically precise metal nanoclusters connect the molecular and bulk worlds, especially those with a core-shell structure where one metal forms the core and another forms the shell\cite{Ferrando_845_2008, Wang_14023_2009, Gawande_7540_2015, Mendes_1158_2020, Eom_8883_2021}. These clusters have become important in catalysis and materials science because their properties can be adjusted, and their surfaces are highly reactive. Transition-metal (TM) clusters exhibit quantum confinement, $d$-band effects, and strong surface effects, making their electronic structures highly sensitive to size and shape\cite{Chakraborty_8208_2017, Castleman_2664_2009}. As a result, both single-metal and bimetallic clusters can display unique energy levels, unexpected magnetism, improved catalytic activity, and unusual optical properties. For example, small icosahedral clusters like \ce{Au13} and \ce{Ag13} can have magnetic moments that do not appear in their bulk forms\cite{Yamamoto_116801_2006, Pereiro_063204_2007}.

When nanoclusters contain so-called ``magic'' numbers of atoms (like \num{13} or \num{55}), they often form stable, closed-shell structures such as the icosahedron (ICO)\cite{Mackay_916_1962, Aiken_1_1999}. This stability comes from their geometry, electronic shell closure, and efficient atomic packing. In addition to their robust structure, these highly symmetric clusters can be tailored to specific electronic and magnetic properties, making them promising for catalysis, sensing, and energy conversion applications\cite{Alonso_637_2000, Baletto_371_2005, Ferrando_845_2008, Fernando_6112_2015}. The possibilities expand even further with bimetallic nanoclusters, such as nano-alloys and core-shell structures, which offer an even wider range of design options. When combined with different metal species, they leverage the metals' interactions to influence adsorption and electronic properties, potentially leading to improved stability, reactivity, and selectivity\cite{Toshima_1179_1998, Ferrando_845_2008, Gawande_7540_2015, Ke_806_2025}. Beyond that, hetero-metallic clusters can have better performance in \ce{CO} oxidation and hydrogen-evolution reactions through cooperative effects\cite{Batista_7431_2019, Wu_78_2019, Kaya_37209_2023}.

Predictive artificial intelligence (PAI) now provides a systematic way to speed up discoveries in this area. High-throughput density functional theory (DFT), an approach that uses DFT calculations to predict the properties of a large number of materials in a short time, combined with algorithmic exploration, has mapped low-energy nanocluster structures at scale, as shown by the Quantum Cluster Database (QCD)\cite{Manna_308_2023, Soleymanibrojeni_2249_2024}. Building on this, transformer models have shown they can combine different types of structural and electronic information using self-attention, leading to reliable and interpretable predictions of material properties\cite{Huang_arxiv_2020}.

\begin{figure*}[!ht]
\centering
\includegraphics[width=0.9\linewidth]{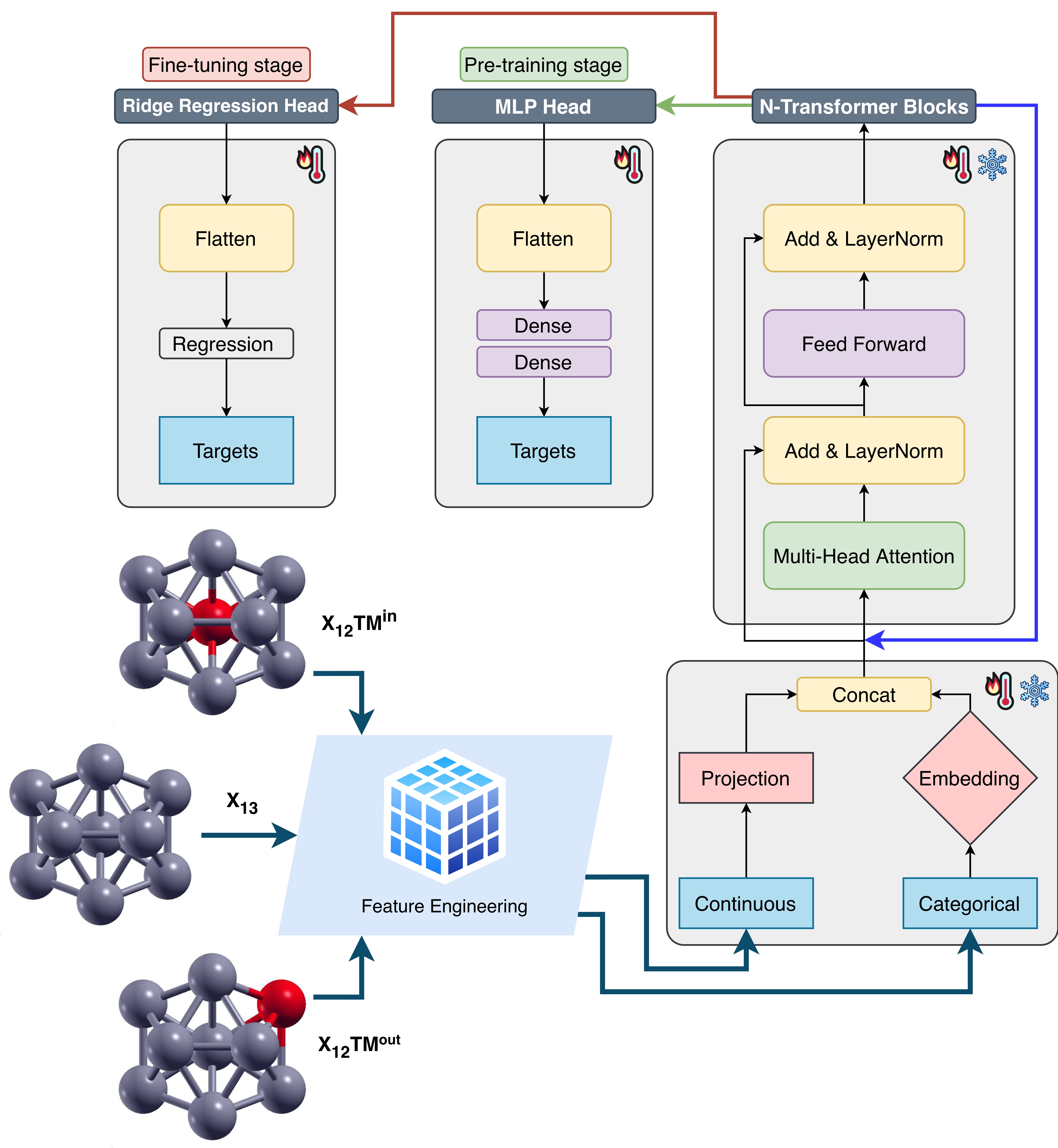}
\caption{Schematic of the dataset and model pipeline. A \num{13}-atom icosahedral cluster \ce{X13} is considered with two single-atom substitution topologies: ($i$) inner-site (core) replacement \ce{X$_{12}$TM$^{in}$} and ($ii$) outer/surface-site replacement \ce{X$_{12}$TM$^{out}$}. Here \ce{X} $\in$ \{\ce{Ti}, \ce{Zr}, \ce{Hf}\} and TM spans the \num{3}$d$--\num{5}$d$ series; $in$ and $out$ denote substitution at the central atom and at one of the \num{12} surface vertices, respectively. Feature engineering produces continuous and categorical descriptors that are projected/embedded and concatenated, then passed through $N$ Transformer blocks (frozen during head-only fine-tuning) to capture cross-feature interactions. A lightweight MLP head (trainable) flattens the representation and predicts the targets, e.g., $E_{form}$ and $\Delta E_{tot}$ for $in$ and $out$ cases.}
\label{fig1}
\end{figure*}

In this paper, as illustrated in Fig.~\ref{fig1}, we focus on \num{13}-atom bimetallic icosahedral clusters composed of \num{12} early-row group-IV hosts (\ce{Ti}, \ce{Zr}, \ce{Hf}) using a single dopant drawn from the \num{3}$d$, \num{4}$d$, and \num{5}$d$ series. Beyond that, we analyze core--shell ($in$) and surface-segregated ($out$) configurations using DFT to provide structural, vibrational, energetic, and electronic characterizations. To generalize these insights, we develop a PAI framework centered on a transformer architecture trained to internalize the physics and chemistry of pure metallic clusters using a curated subset of low-energy structures from QCD\cite{Manna_308_2023}. After that, we fine-tuned this pretrained model on a small, high-quality DFT dataset of bimetallic clusters, enabling accurate predictions of binding and formation energies and the $in$/$out$ configuration preference. The attention structure affords interpretability, exposing which atomic environments, shells, and electronic descriptors drive stability trends, while offering design rules aligned with known $d$-band and size/geometry effects.

Our workflow follows the FAIR (Findable, Accessible, Interoperable, Reusable) and TRUE (Transparent, Reproducible, Usable by Others, and Extensible) data principles\cite{Rego_2022, Schaarschmidt_2102638_2021, Bastos_2025}. Following these ideas, we make sure that our data and methods are open, reproducible, and easy for others to use or build on. We share the workflow in a public repository so anyone can find the inputs, models, and outputs and work with the latest version. We also keep track of data sources and uncertainties and provide clear instructions for reuse\cite{Rego_XXXX_2025, Bekemeier_XXXX_2025, Dalmedico_e202400118_2024}. This setup offers a flexible, reproducible framework that combines advanced DFT methods with interpretable transformers. It can transfer knowledge from pre-trained to fine-tuned models, needing less data than a fully data-driven method. This approach allows us to better understand how stability, distortion, and reactivity interact in bimetallic nanoclusters, helping us design new catalysts and materials for energy conversion.

\section{Results and Discussions}

\subsection{DFT-based trends} 

\paragraph{Structural stability:}  The vibrational frequency analysis, as shown in Supporting Information (\SM~Fig.~S10), shows that all optimized \ce{X$_{12}$TM$^{in}$} and \ce{X$_{12}$TM$^{out}$} nanoclusters are stable, as they are local minima on the potential energy surface with only real, positive frequencies. This confirms their mechanical stability. Interestingly, the $out$ configurations often have slightly higher vibrational frequencies for specific modes. This happens because these structures are more geometrically distorted than the more symmetric $in$ (core-shell) arrangements, leading to stiffer local bonding in some less symmetric cases.

\begin{figure*}[!ht]
\centering
\includegraphics[width=0.85\linewidth]{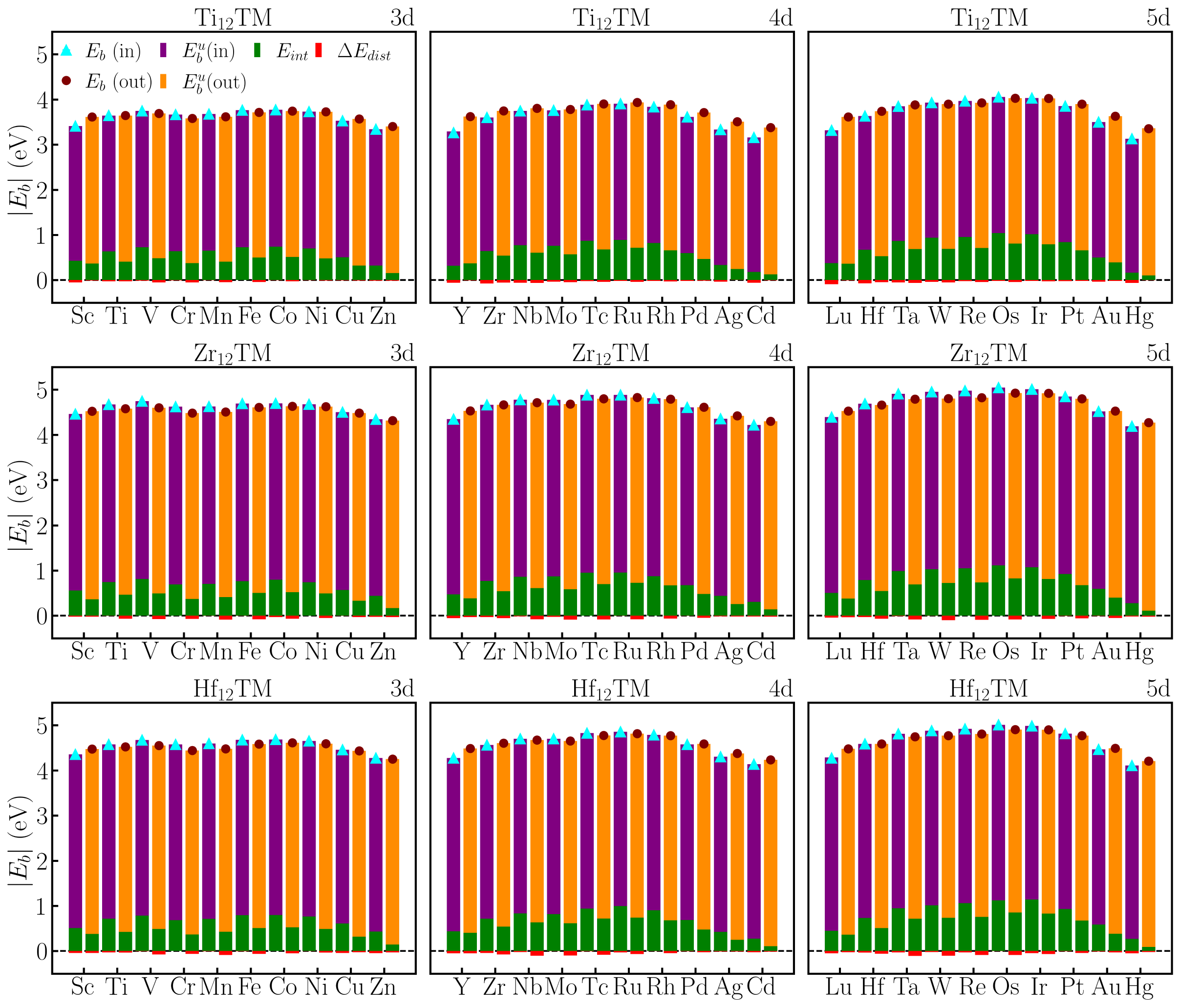}
\caption{The binding energy modulus is shown as function of the atomic number for \ce{X$_{12}$TM$^{in}$} (triangle) and \ce{X$_{12}$TM$^{out}$} (circle) (\ce{X} = \ce{Ti}, \ce{Zr}, \ce{Hf}) nanoclusters. The energy decomposition presents the binding energy of \ce{X$_{12}$} ICO-derived structure ($E_b^u$), the interaction energy between \ce{X$_{12}$} and TM in the equilibrium geometry of \ce{X$_{12}$TM$^{in}$} or \ce{X$_{12}$TM$^{out}$} ($E_{int}$), and the distortion energy occasioned by interaction between TM species and \ce{X$_{12}$} structure (${\Delta}E_{dist}$).}
\label{fig2}
\end{figure*}

The computed binding energies shown in Fig.~\ref{fig2} confirm that all binary nanoclusters, regardless of the TM position ($in$ or $out$), are thermodynamically stable, exhibiting negative values. This trend is consistent with the strong metallic bonding characteristic of group IV metals (\ce{Ti}, \ce{Zr}, \ce{Hf}) and their interactions with other TM\cite{Harrison_2012}. As shown, we observe a negative parabolic trend in binding energy modulus as a function of the TM atomic number, consistent with the classical $d$-band filling model, i.e., a behavior analogous to that observed in bulk TM systems and small metallic clusters\cite{Harrison_2012, Piotrowski_155446_2010, Chaves_15484_2017}. This arises from the systematic filling of the $d$-orbitals across the transition series. For early TM elements (\ce{Sc} to \ce{Cr}), the progressive filling of bonding-type $d$-states increases the binding energy modulus, enhancing the stability near the center of the series (\ce{Mn}, \ce{Fe}, \ce{Co}), the occupation of non-bonding and anti-bonding states begins to counteract the bonding gains, leading to a peak in stability. Toward the end of the series (\ce{Zn}, \ce{Cd}, \ce{Hg}), the full occupation of $d$-shells ($d^{10}$ configuration) reduces the availability of bonding states, the interactions become predominantly repulsive due to filled anti-bonding character, sharply decreasing the binding energy modulus and contributing minimally to metallic bonding, as well as inducing anti-bonding interactions due to Pauli repulsion. 

The decomposition of binding energy in Fig.~\ref{fig2} into its fundamental components, $E_{int}$ and $\Delta E_{dist}$, reveals a competitive interplay governing stability. The core-shell ($in$) configurations generally benefit from enhanced $E_{int}$ compared to their $out$ counterparts, which is attributed to the maximized coordination of the central TM atom, since it interacts with \num{12} nearest neighbors in the icosahedral cage. However, these configurations are penalized by geometric strain when there is a significant mismatch in size or electronic character between the dopant and host atoms. In contrast, the $out$ configurations experience lower interaction energies because, when the dopant is surface-coordinated, the TM atom interacts with a reduced number of ligands, typically \num{6} to \num{7}, resulting in lower interaction energy and greater exposure to surface-induced distortions. 

Distortion energy plays a crucial role in modulating stability. The nanoclusters with lower binding energy modulus are more likely to have higher distortion energies, particularly the late TMs (\ce{Zn}, \ce{Cd}, \ce{Hg}), where mismatch of electronic character and atomic sizes between dopant and host (\ce{Ti}, \ce{Zr}, \ce{Hf}) causes geometric frustration. Interestingly, despite the larger distortion energies observed in $out$ nanoclusters, the interaction energies are, in most cases, sufficient to compensate for them, preserving the overall thermodynamic stability. However, in specific systems where the distortion penalty becomes comparable to or exceeds the interaction gain, mainly with larger or electronically inert dopants, the $out$ configuration becomes less favorable, as quantified by the relative total energy.

\begin{figure*}[!ht]
\centering
\includegraphics[width=0.5\linewidth]{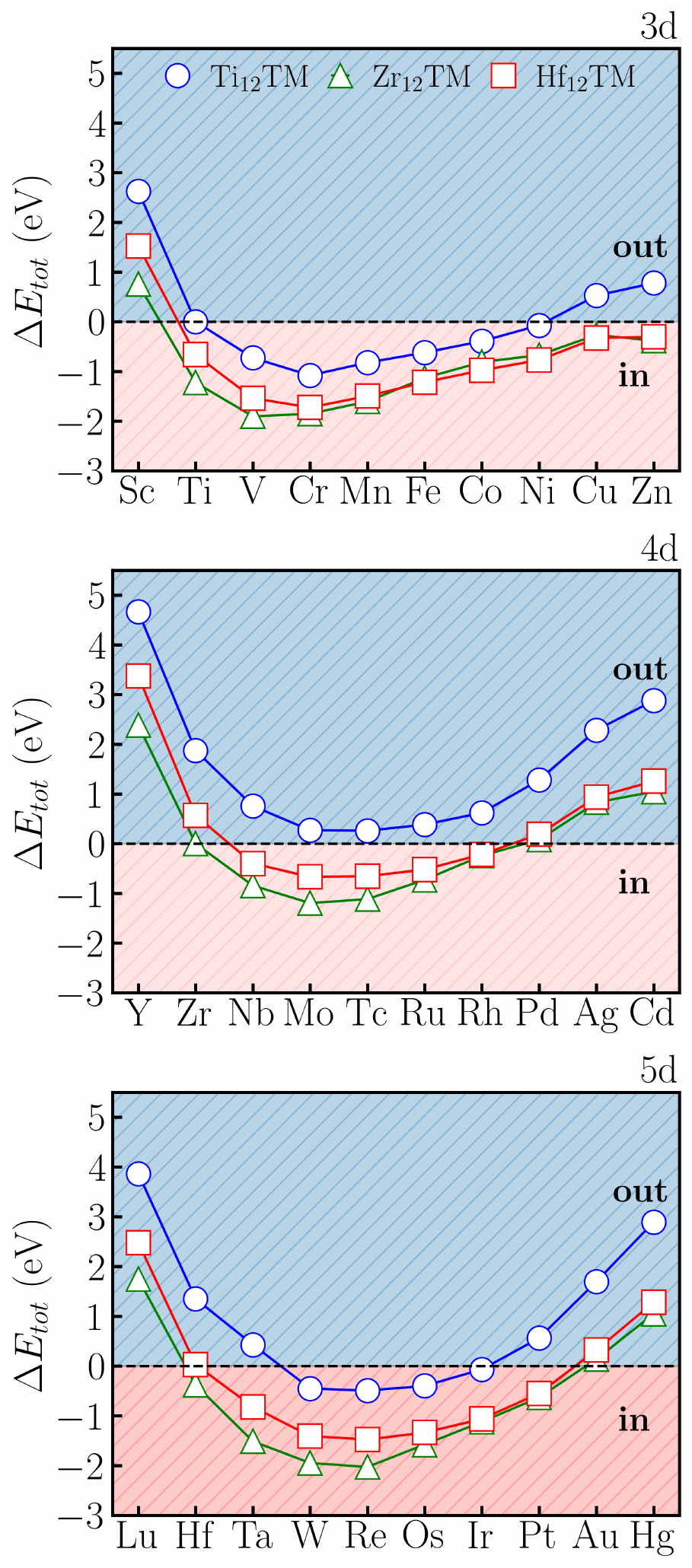}
\caption{Relative total energy ($\Delta E_{tot}$) as a function of atomic number for \ce{Ti$_{12}$TM} (blue circles), \ce{Zr$_{12}$TM} (green triangles), and \ce{Hf$_{12}$TM} (red squares) nanoclusters. Negative values indicate energetic stabilization of $in$ configurations, while positive values indicate stabilization of $out$ configurations.}
\label{fig3}
\end{figure*}

Thus, a more direct form to indicate the TM energetic preference in $in$ and $out$ configurations is obtained by subtracting the binding energies of the respective $in$ and $out$ configurations of \ce{X$_{12}$TM}, resulting in the $\Delta E_{tot}$, as shown in Fig.~\ref{fig3}. The $\Delta E_{tot}$ profile quantitatively captures the energetic preference between $in$ and $out$ configurations, since negative values indicates a thermodynamic preference for the core-shell structure, predominantly observed for mid-row TMs (except for \num{4}$d$ species in \ce{Ti$_{12}$TM}), where the combination of bonding strength and favorable electronic hybridization maximizes stability. In contrast, positive values of $\Delta E_{tot}$ are characteristic of early and late TMs, where geometric strain or electronic inertness favors the dopant occupying surface ($out$) positions. The formation energies $E_{form}$ as displayed in \SM~Fig.~S11 reinforce these trends, confirming that the most stable configurations are those with minimized distortion and maximized bonding synergy between the host (\ce{Ti}, \ce{Zr}, \ce{Hf}) and the dopant.

For \ce{Fe$_{12}$TM} nanoclusters, we observe that energetically they exhibit trends analogous to those observed for the group IV hosts (\ce{Ti}, \ce{Zr}, and \ce{Hf}). The binding energies are consistently negative across the transition series, indicating thermodynamic stability regardless of the dopant position. A distinctive feature in \ce{Fe}-based nanoclusters is the influence of \ce{Fe} atomic radius and significant magnetic character. This effect leads to a larger relative distortion when dopants occupy the central site, particularly for heavier and larger TMs (e.g., \ce{Au}, \ce{Hg}). As a result, the distortion energy penalty increases for $in$ configurations, making $out$ arrangements energetically more competitive in selected cases, as clearly shown in the relative stability plots (Figs. ~S12 and S13). This is due to size mismatch and magnetic effects.

\begin{figure*}[!ht]
\centering
\includegraphics[width=0.95\linewidth]{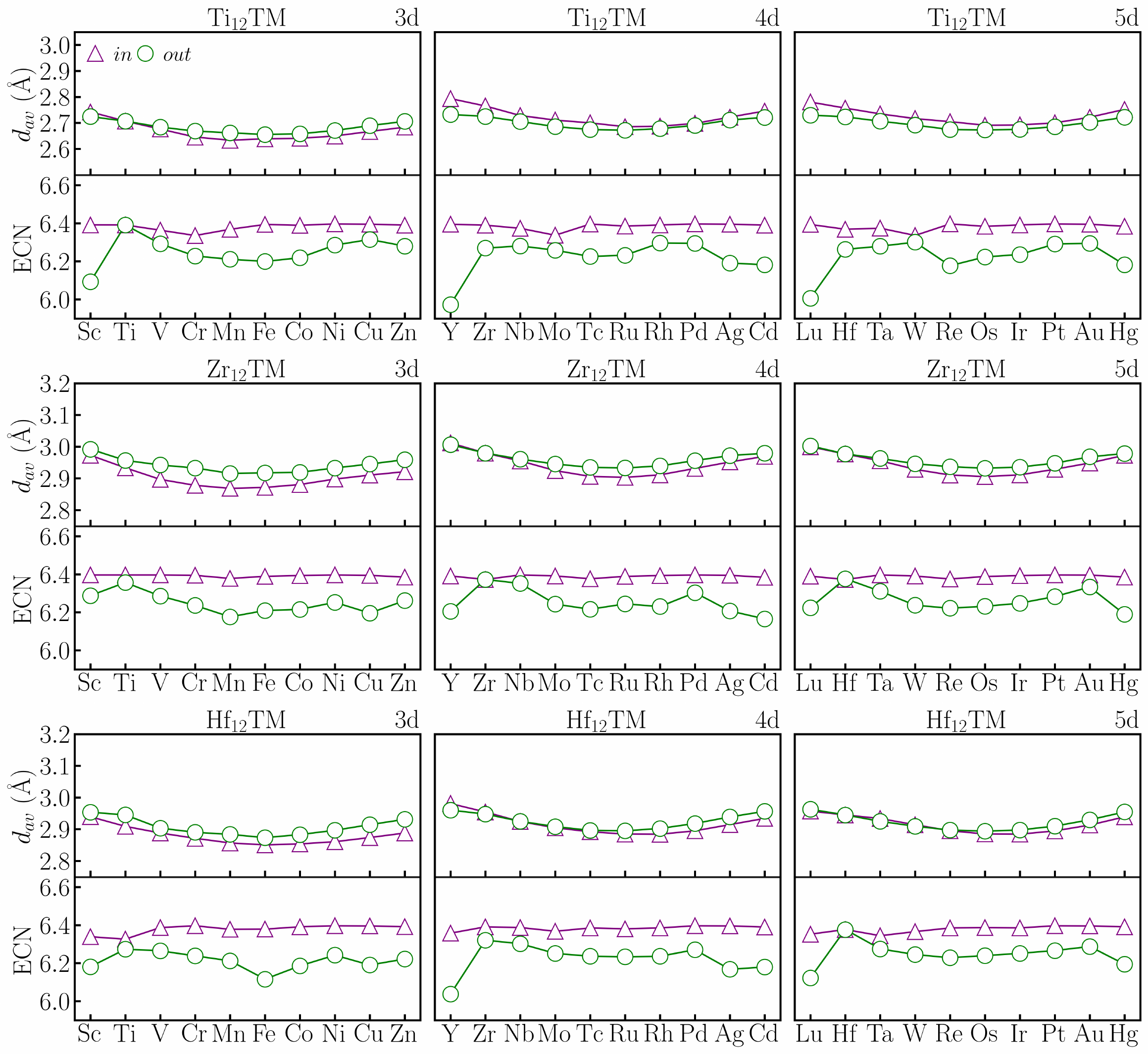}
\caption{Structural properties: the average bond lengths ($d_{\text{av}}$) and the effective coordination number (ECN) as function of the TM atomic number for \ce{X$_{12}$TM$^{in}$} (triangle) and \ce{X$_{12}$TM$^{out}$} (circle) (\ce{X} = \ce{Ti}, \ce{Zr}, \ce{Hf}) nanoclusters.}
\label{fig4}
\end{figure*}

\paragraph{Geometric properties:} The substitution of a TM atom as $in$ or $out$ configurations has a profound effect on the geometric and electronic structure of the nanoclusters. To capture this primarily from a geometric perspective, we have evaluated two key metrics, $d_{\text{av}}$ and ECN, shown in Fig.~\ref{fig4}. Core-shell structures always consist of smaller $d_{\text{av}}$ values and higher ECN, signifying a closer-together bonding environment and a more symmetric arrangement inherent in the icosahedral cage. The high coordination of the central TM atom (ECN = \num{12}) leads to stronger bonding, which shortens bond distances. On the other hand, $out$ configurations exhibit $d_{\text{av}}$ elongation and a significant reduction in ECN. This is attributed to two factors: surface localization of the dopant, which inherently reduces the number of nearest neighbors, and local relaxation effects, as the nanocluster adapts its geometry to accommodate size and electronic mismatch. For instance, late TMs such as \ce{Cd}, \ce{Hg}, and \ce{Au} exhibit the most pronounced deviations, characterized by bond length expansions exceeding \SI{5}{\percent} relative to their $in$ counterparts.

These structural metrics correlate directly with the energetic descriptors. For example, in $out$ systems, structural flexibility is directly associated with weaker binding energies and higher distortion penalties. Systems with larger $d_{\text{av}}$ and lower ECN tend to show higher $\Delta E_{tot}$ and less negative binding energies, particularly in the $out$ configurations, supporting the classical principle that under-coordinated sites are structurally less stable but potentially more reactive. In this context, as shown in Fig.~\ref{fig4}, the ECN is a quantitative descriptor of the local atomic environment between the two cases ($in$ or $out$). In $in$ configurations, the TM dopant achieves the maximum ECN, which is associated with more efficient bonding and higher stability. In $out$ configurations, the ECN decreases significantly, typically below \num{7}, depending on the specific TM and the host nanocluster, which indicates under-coordination and higher chemical reactivity. This behavior aligns with general trends in metallic bonding: atoms tend to maximize their coordination to lower the system's energy. However, if the TM is placed outside the cage, steric hindrance and size mismatch can prevent ideal coordination, leading to a metastable yet locally stable structure.

For \ce{Fe$_{12}$TM} nanoclusters, we observe that, structurally, $d_{\text{av}}$ and ECN (Fig.~S14) reflect systematic differences between $in$ and $out$ configurations. The $in$ systems possess shorter $d_{\text{av}}$ and higher ECN values, reflecting the close coordination environment of the center \ce{TM} atom. The $out$ structures show a sharp reduction in ECN (typically below \num{7}), which reflects the under-coordination and greater surface exposure (accessibility) trend.

\paragraph{Electronic properties:} The energetic trends are intimately connected to the electronic structure of the binary nanoclusters, i.e., a detailed analysis of the electronic properties, as shown in \SM~Figs. ~S15 and ~S16 reveal a profound correlation between electronic structure and thermodynamic stability. In core-shell configurations, the different central TM atom experiences a ligand field analogous to a highly symmetric crystal field, splitting the $d$-orbitals into bonding and anti-bonding combinations with the surrounding shell. The degree of splitting and orbital hybridization depends on the $d$-electron count of the dopant, e.g., early TMs contribute to strong $\sigma$ and $\pi$ bonding interactions, enhancing stability; mid-series TMs reach maximum stabilization due to half-filled or nearly half-filled $d$-subshells, favoring exchange interactions; and late TMs exhibit filled $d$-states with minimal bonding contribution, resulting in weaker metal--metal interactions dominated by Pauli repulsion and size effects.

TMs-centered systems ($in$) register systematically lower $\epsilon_d$ values as displayed in \SM~Fig. ~S15. This shift down signals stronger hybridization between the TM $d$-orbitals and the metallic cage, stabilizing bonding states and lowering the energy. In $out$ configurations, $\epsilon_d$ moves up, signaling weaker hybridization and a greater fraction of unsaturated or dangling electronic states. On the other hand, the HOMO--LUMO gap (\SM~Fig. ~S16) reflects electronic stability and potential chemical reactivity. Core-shell nanoclusters have larger gaps, consistent with an electronically more closed-shell, less reactive system. In contrast, $out$ nanoclusters have systematically smaller gaps, particularly for mid-row and late-row TMs. This trend suggests that $out$-nanoclusters could exhibit enhanced chemical reactivity, making them promising candidates for catalytic applications where active sites at undercoordinated positions are desirable. We found that there is an anti-correlation between $\epsilon_d$ and the HOMO--LUMO gap, as $\epsilon_d$ increases (i.e., approaches the Fermi level), the HOMO--LUMO gap narrows, in agreement with the expected behavior from the $d$-band center theory\cite{Hammer_211_1995, Hammer_71_2000}, which links the position of $\epsilon_d$ to adsorption strength and catalytic activity. 

For \ce{Fe$_{12}$TM} nanoclusters, we observe that from the electronic perspective, the HOMO-LUMO gap (\SM~Figs. S17 and S18) is typically narrower for $out$ configurations, suggesting higher chemical reactivity and catalytic activity. Variations in electronic gaps also support the established understanding that $in$ configurations, with stronger ligand fields and higher symmetry, lead to electronic delocalization and increased stability. Therefore, it is important to note that these electronic descriptors -- $\epsilon_d$ and the HOMO--LUMO gap -- provide the physics-grounded signals from which we have used for our PAI supervision. The nearly \num{3000} set of relaxed unary nanoclusters from the QCD and the \num{240} bimetallics DFT computations form a corpus that serves as the basis for the pre-training and fine-tuning stages discussed next.

\subsection{Physics-Grounded Signals to Transformers Performance} 

After the hyperparameter search described in Methods, we selected the final model by comparing its complexity and performance during finetuning. The chosen FTTransformer has \num{6} transformer blocks with \num{8} attention heads, an embedding dimension of \num{16}, and a feed-forward width of \num{128}. Transformer blocks use swish, and the MLP head uses gelu activation with hidden-size factors (\num{2}, \num{1}) -- the first hidden layer has twice the flattened transformer output, and the second equals it. During fine-tuning, the backbone transformer and tokenizer remain frozen. More information can be found in the SI and in the GitHub \href{https://github.com/KIT-Workflows/Nanocluster_Transformers}{repository}.

The finetuning was performed using robust Ridge regression. The original data is highly non-linear and complex. The transformer's intricate architecture (attention mechanisms, residual connections, multiple layers) learns this complexity (Fig. S19). It combines relevant features from the input in a way that makes the underlying patterns more explicit and easier to discern. In this new space of transformed representation (embeddings), the features are structured such that the target variable ($E_{form}^{in}$ and $E_{form}^{out}$) can be approximated well by a linear combination of these features. Therefore, we leverage the transformer architecture to simplify the fine-tuning stage, making it much more robust, faster, and numerically stable.

\begin{figure*}[!ht]
\centering
\makebox[0pt][r]{(a)\hspace{1em}}%
\includegraphics[width=0.65\textwidth]{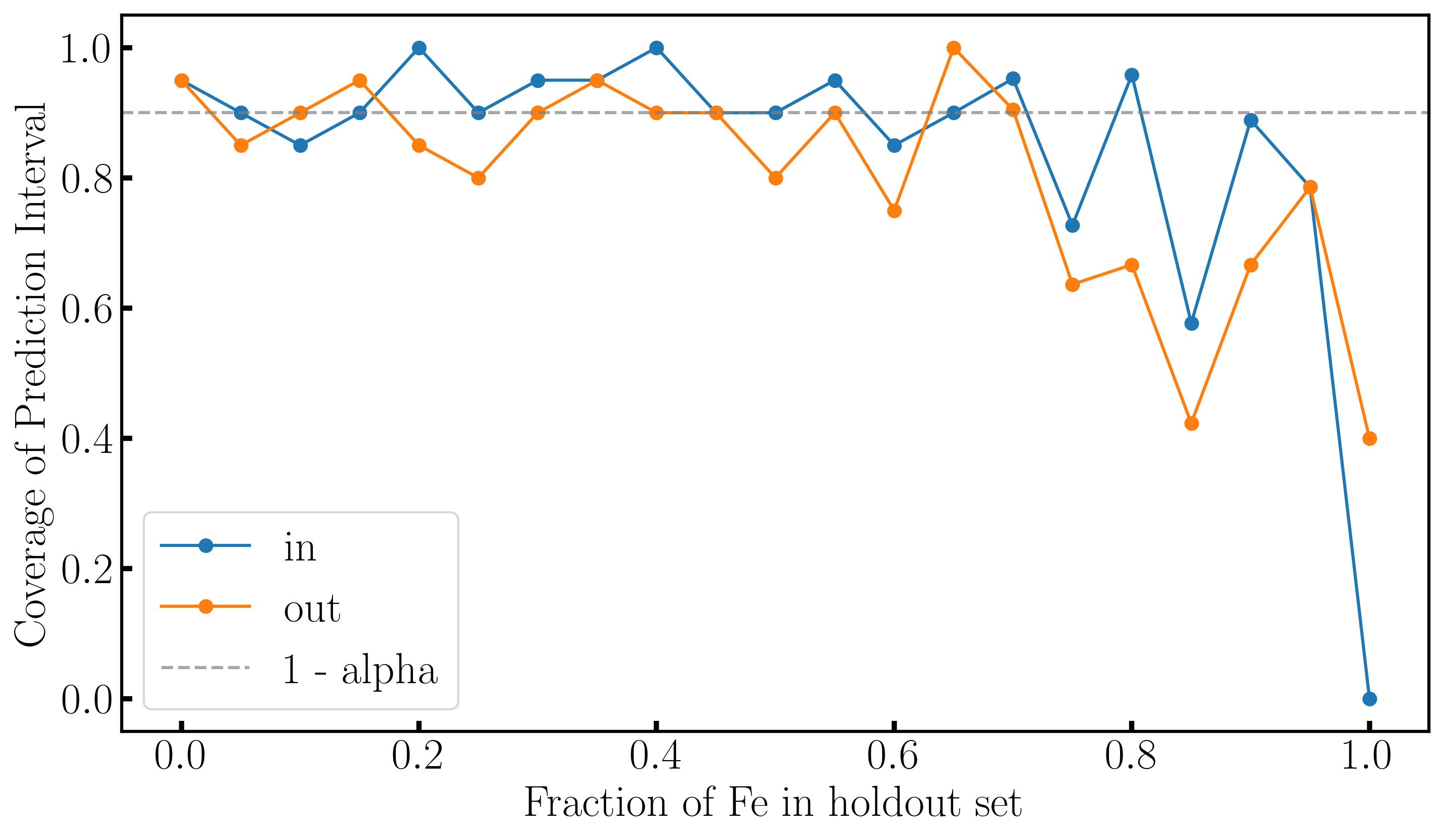}
\\[0.1em]
\makebox[0pt][r]{(b)\hspace{1em}}%
\includegraphics[width=0.65\textwidth]{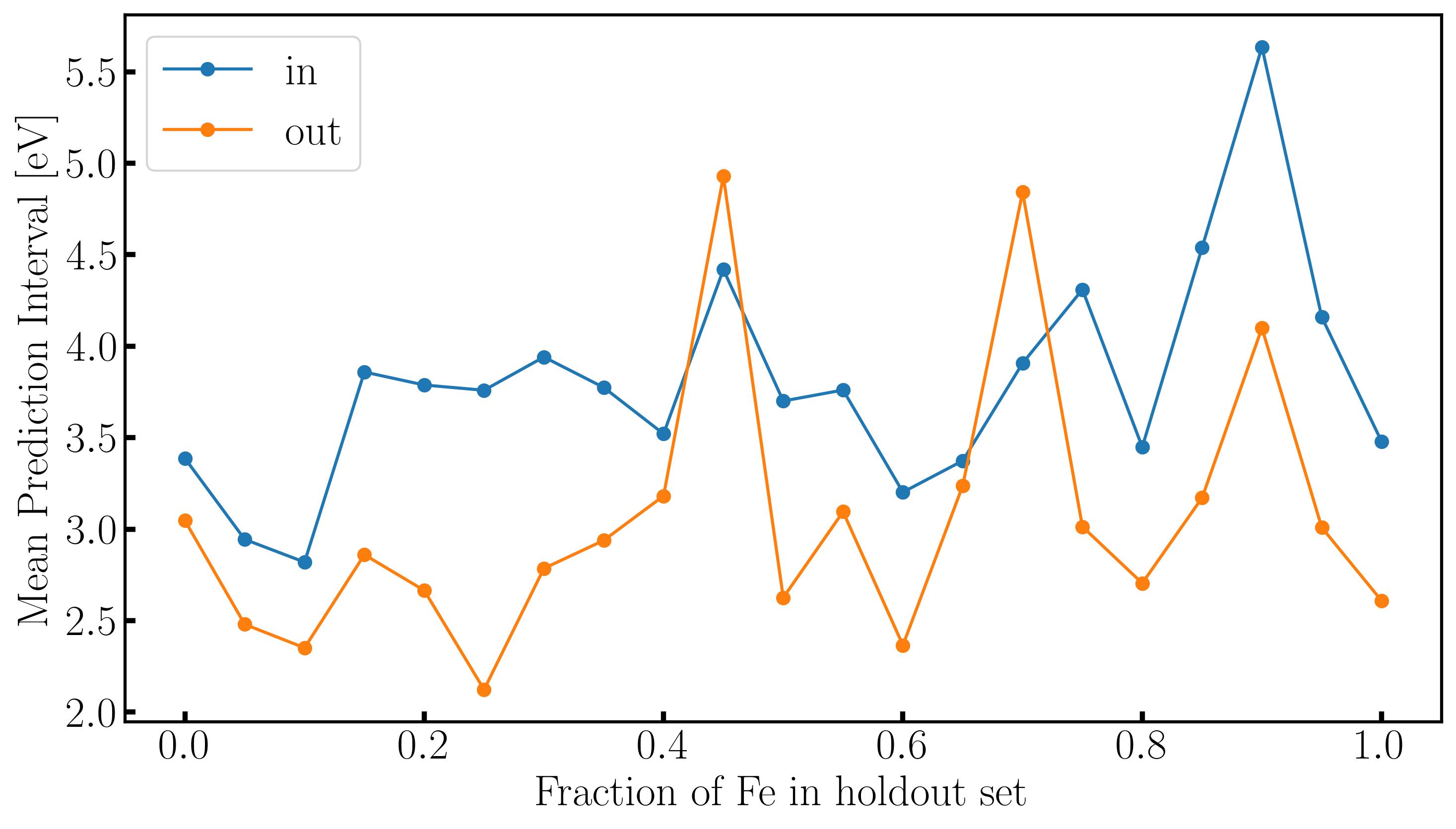}
\\[0.1em]
\makebox[0pt][r]{(c)\hspace{1em}}%
\includegraphics[width=0.65\textwidth]{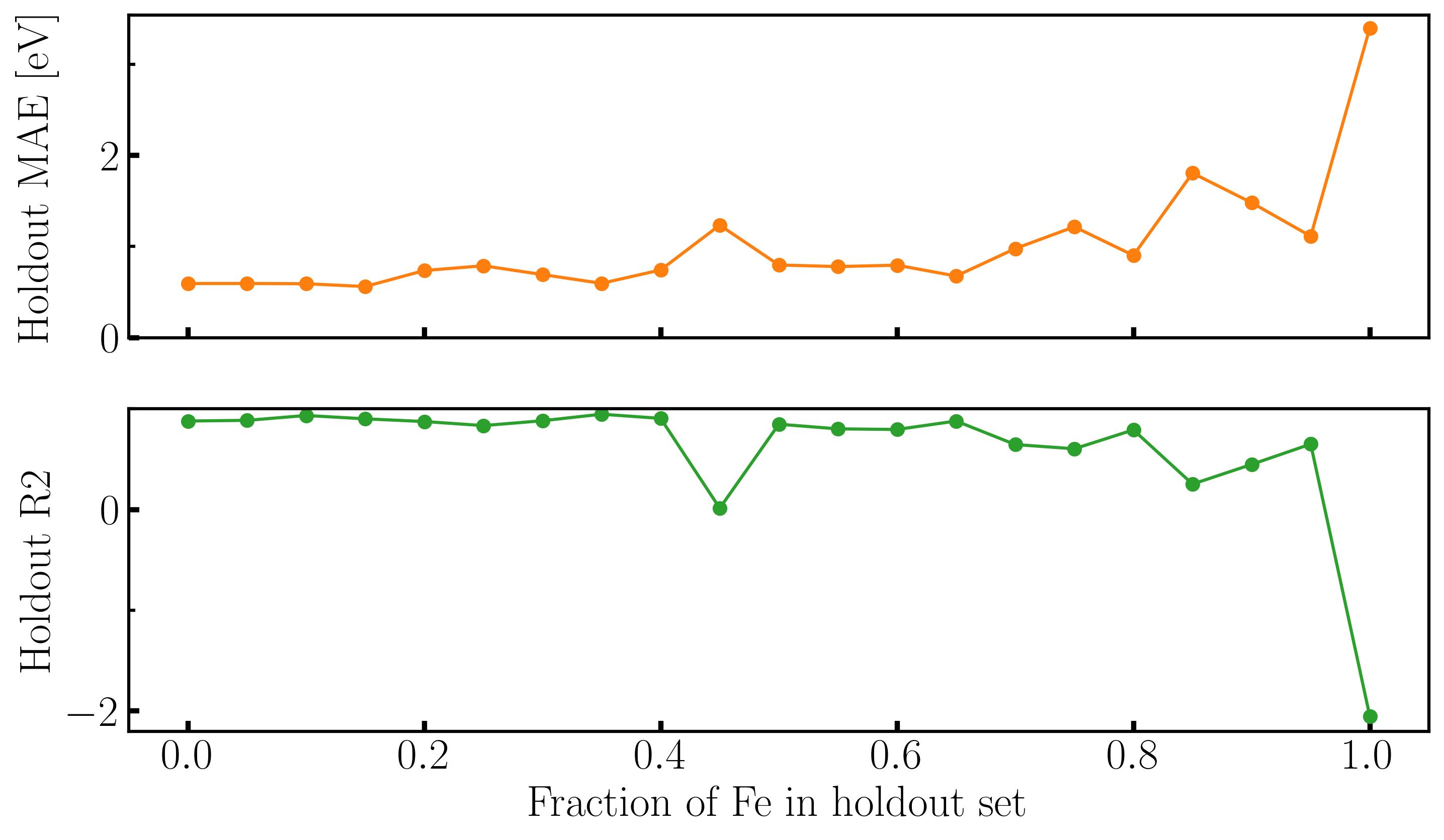}
\caption{Fraction of the total \ce{Fe} host atoms ($X_{12}$) in holdout set compared against the coverage and width of prediction intervals, as well as MAE and $R^2$ score. This fraction changes between \num{0.0} and \num{1.0}. When \num{0.0} that means all \ce{Fe} host atoms are in the training set and the holdout set does not include any entries with \ce{Fe} as host atom instead with other elements, when \num{1.0} the opposite holds. The minimum size of the holdout set is 20 and the maximum is 30. At $x =0.95$ which is insertion of only \num{2} entries containing \ce{Fe} host atoms in training set, all metrics improve, at $x=0.65$ and lower $x$, we can see a stable trend starts. (a) The coverage of prediction intervals, $\alpha=0.1$, (b) Mean prediction intervals for both \textit{in} and \textit{out} configurations, (c) MAE for both targets in the holdout set, (d) $R^2$ score is the regression score for both targets.}
\label{fig:scan}
\end{figure*}

\begin{figure*}[!ht]
\centering
\subfigure{%
  \makebox[0pt][r]{(a)\hspace{1em}}%
  \includegraphics[width=0.7\linewidth]{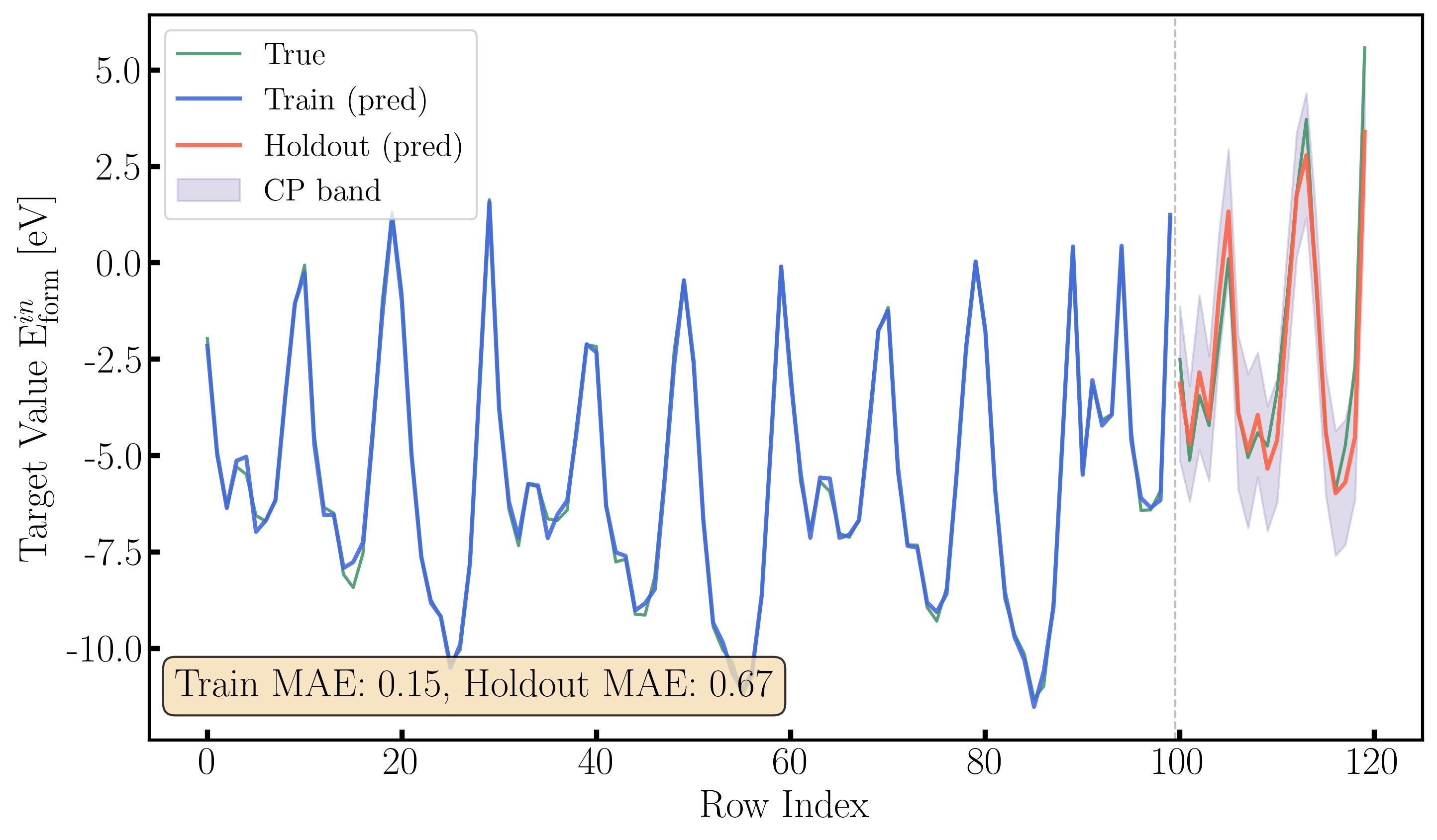}%
}\hfill
\subfigure{%
  \makebox[0pt][r]{(b)\hspace{1em}}%
  \includegraphics[width=0.7\linewidth]{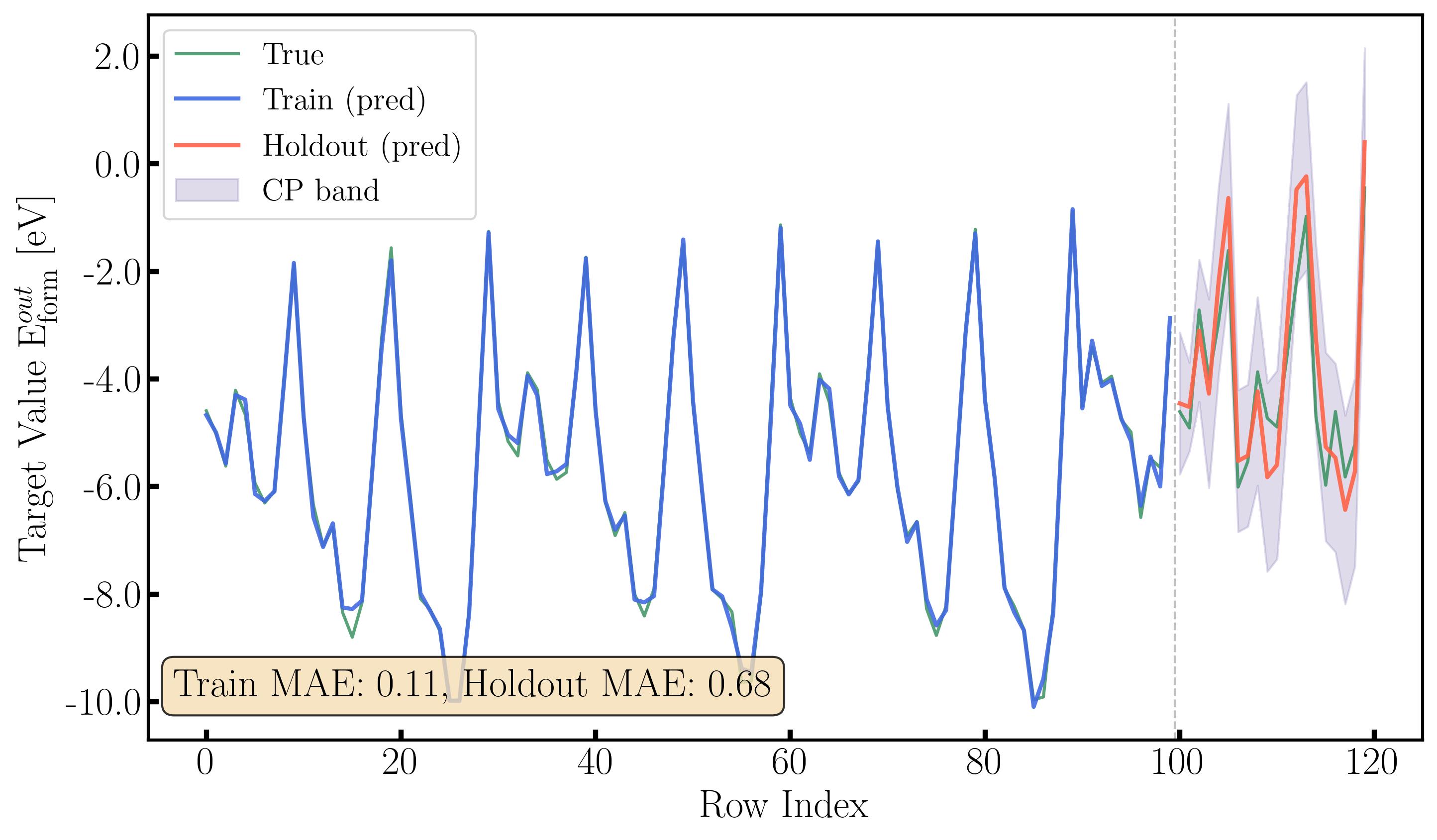}%
}
\caption{Fine-tuning stage with \num{5}-fold CV+, and conformal prediction intervals with $\alpha=0.1$. Bands denote calibrated prediction intervals. For (a) \textit{in} configuration the mean width of prediction intervals is \SI{3.37}{\electronvolt}, coverage \num{0.9}, and holdout MAE \SI{0.67}{\electronvolt}, and for (b) \textit{out} the mean width of prediction intervals is \SI{3.24}{\electronvolt}, coverages \num{1.0}, and holdout MAE \SI{0.68}{\electronvolt}. These results belong to the model trained at $x=0.65$ "Fraction of Fe in holdout set" meaning with \num{10} \ce{Fe} host atoms in the training of the fine-tuned model, and the rest of \num{20} \ce{Fe} host atoms in the holdout set.}
\label{fig:ml_cp_timeline}
\end{figure*}

In this regard, we created different splits of fine-tuning sets. The total \num{120} rows of fine-tuning set is made of \num{30} rows with \ce{Fe} as host atom ($X_{12}$) and the rest of \num{90} have \ce{Ti}, \ce{Zr}, \ce{Hf} as host. By incrementally introducing host \ce{Fe} in the training set of the finetuning stage, we observed the behavior of errors and prediction intervals. Fig.~\ref{fig:scan} Shows when the fraction of host \ce{Fe} in holdout set is full (\num{1.0}) meaning all host \ce{Fe} entries are kept in holdout set and the fine-tuned model was not exposed to any of those entries, the coverage of prediction interval for \textbf{in} and \textbf{out} configurations are \num{0.0} and \num{0.4}, respectively, the MAE \SI{3.4}{\electronvolt}, and $R^2$ score (coefficient of determination) \num{-2}. But after introducing only two rows of host atom \ce{Fe} (fraction of Fe in holdout set = 0.95, 0.05 in training set, $\lceil0.05 \times 30\rceil =2$) in training of the fine-tuned model, the coverage jumps to \num{0.8} and MAE reduced to \num{1.1} and $R^2$ score improved to \num{0.7}. Several factors, including the size of the finetuning set, influence the mean width of prediction intervals. With a limited fine-tuning set, the prediction intervals are kept conservatively large to ensure coverage.
The trend shows a clear, rapid stability of model performance with a small amount of fine-tuning data. We observe that when the fraction of \ce{Fe} in the holdout set is $x = 0.65$ (model fine-tuned with 10 \ce{Fe} host entries), the model's performance reaches a level of stability that can be considered representative. We picked this model to show its response details during fine-tuning and evaluation with a holdout set in Fig.~\ref{fig:ml_cp_timeline}.  The results for both targets of $E_{form}^{in}$ and $E_{form}^{out}$ respectively are MAE of \num{0.15} and \num{0.11} for training step, and MAE of \num{0.67} and \num{0.68} on holdout set, the average prediction intervals are \num{3.37} and \SI{3.24}{\electronvolt}, and the coverage of \num{0.9} and \num{1.0}.

\begin{figure*}[!ht]
\centering
\subfigure{%
  \begin{tabular}{@{}r@{}}
    (a) \\
    \includegraphics[width=0.49\linewidth]{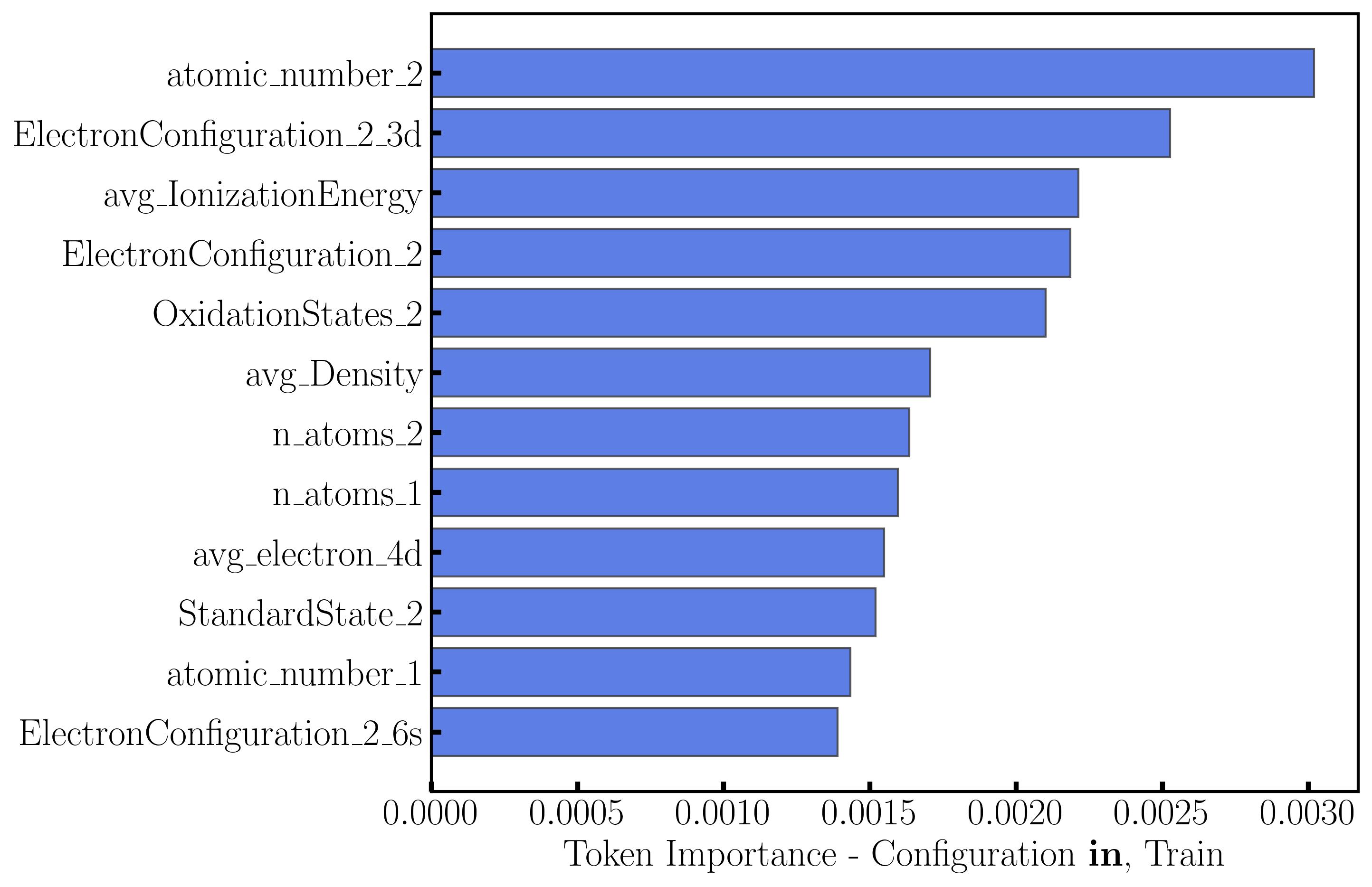}
  \end{tabular}%
}
\hfill
\subfigure{%
  \begin{tabular}{@{}r@{}}
    (b) \\
    \includegraphics[width=0.49\linewidth]{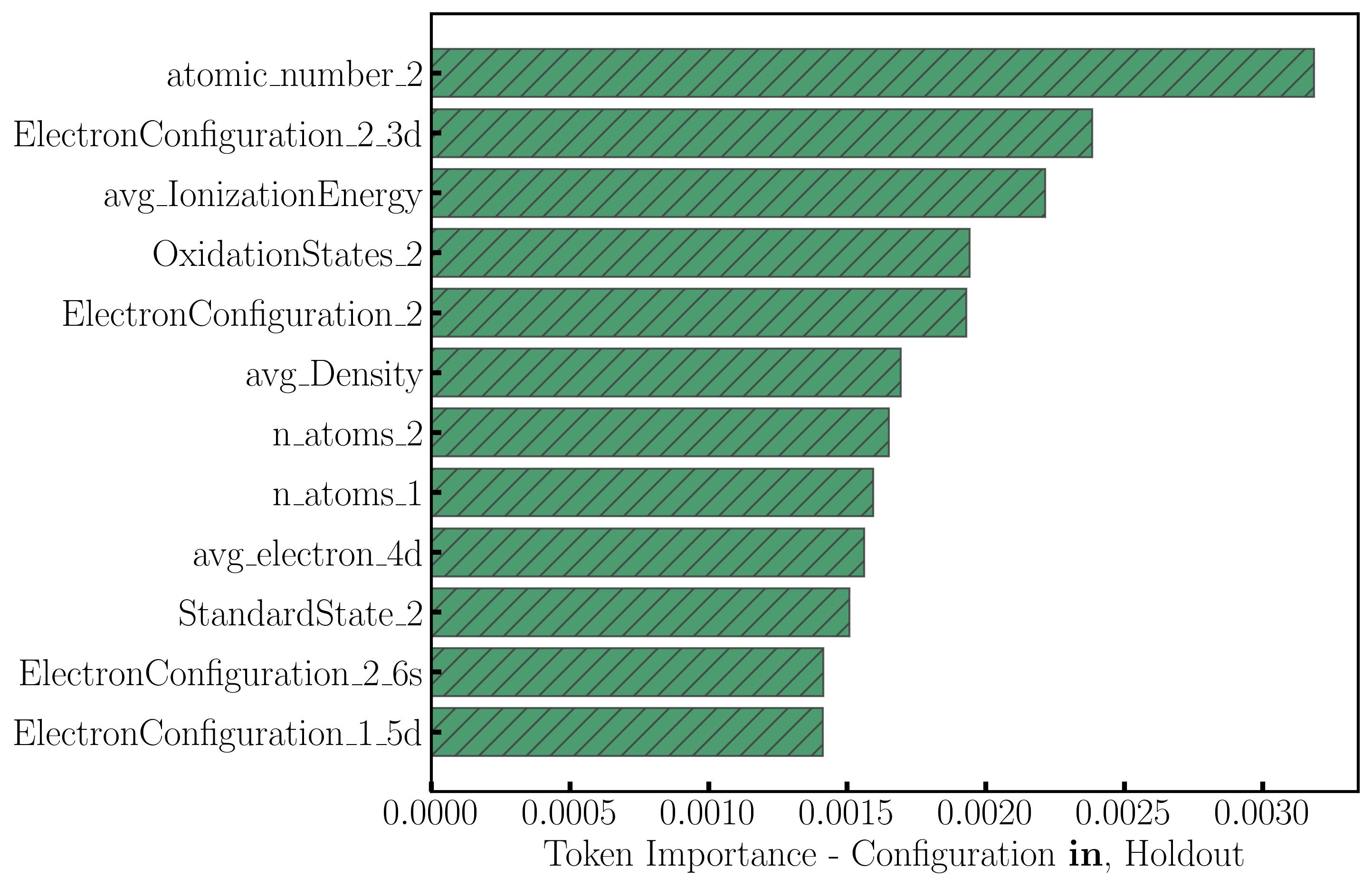}
  \end{tabular}%
}
\\
\subfigure{%
  \begin{tabular}{@{}r@{}}
    (c) \\
    \includegraphics[width=0.49\linewidth]{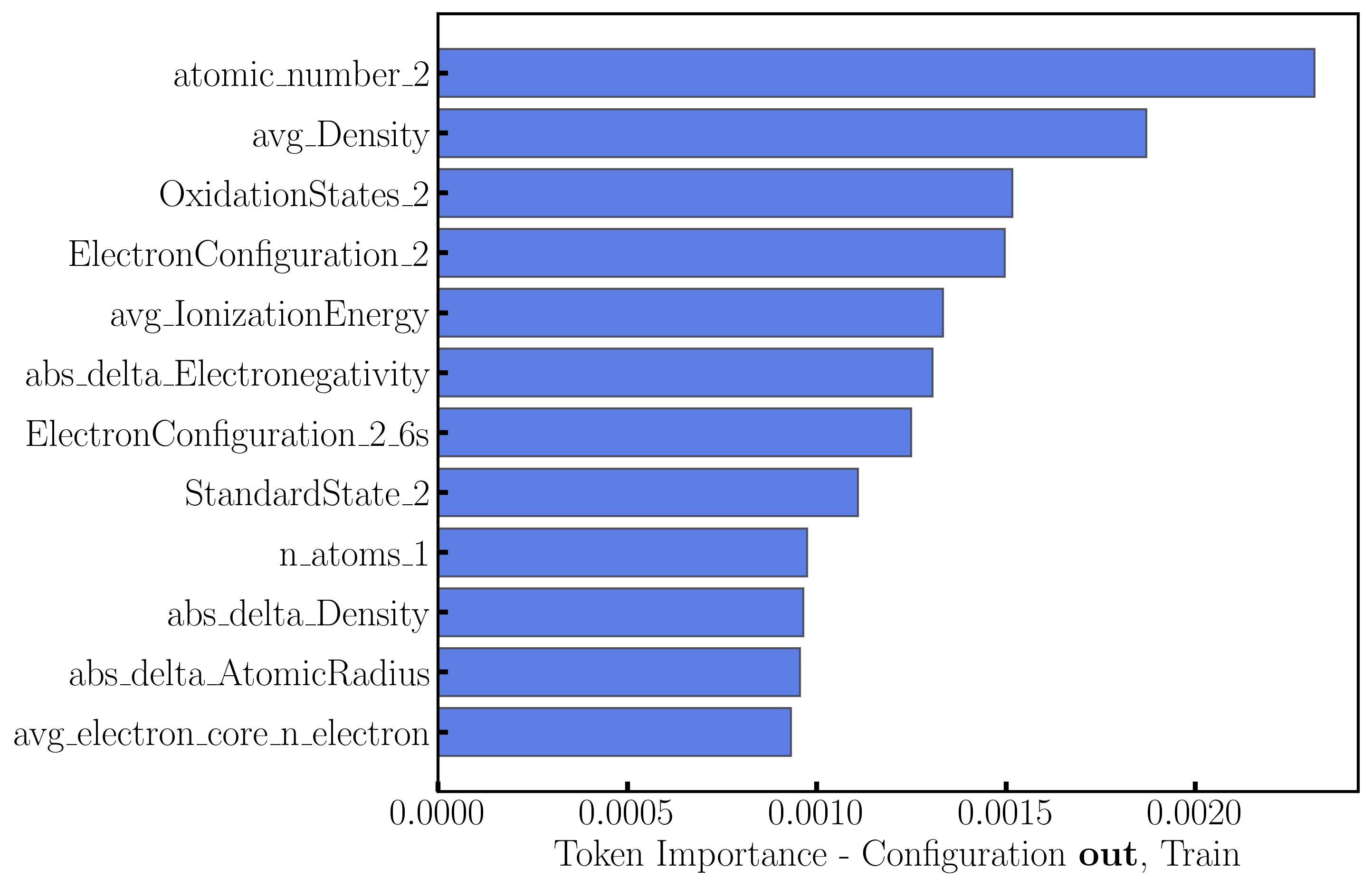}
  \end{tabular}%
}
\hfill
\subfigure{%
  \begin{tabular}{@{}r@{}}
    (d) \\
    \includegraphics[width=0.49\linewidth]{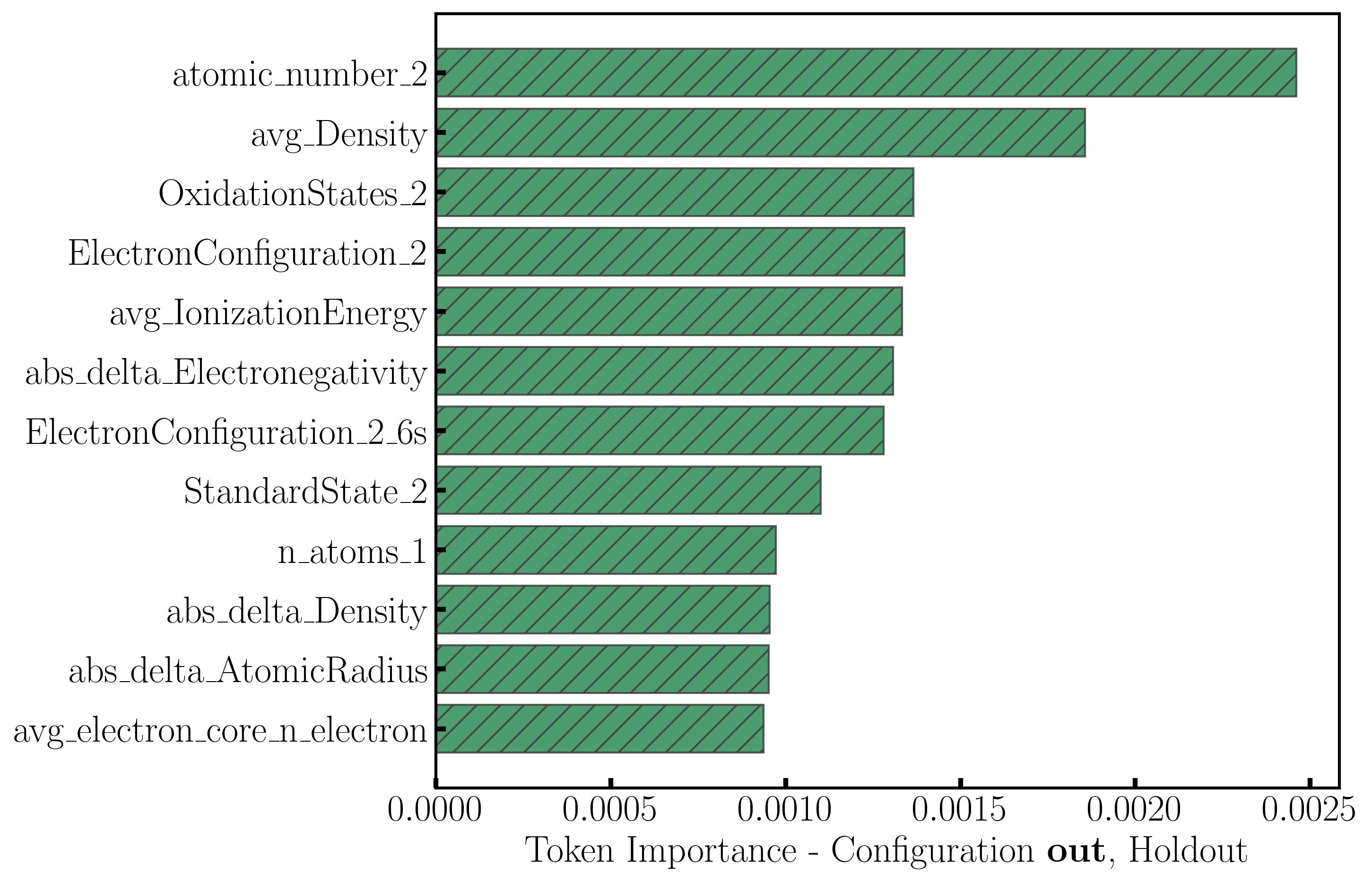}
  \end{tabular}%
}
\caption{Token importance for both configurations, comparing fine-tune and holdout. Dominant contributors include electron-configuration descriptors and periodic-table properties, as well as engineered features (absolute differences and averages). The stability of ranking across splits supports physically meaningful representations and transferred learning. (a,b) Training and holdout sets for \textit{in} configuration, and (c,d) Training and holdout sets for \textit{out} configuration. Similar token importance ranking in both the training and holdout sets, as well as distributed importance across different features, are indications of a reliably fine-tuned model.}
\label{fig:ml_shap}
\end{figure*}

To gain a clearer view of the model's underlying behavior, we performed token importance analysis at both training and evaluation, using a holdout set for the fine-tuning stage (Fig.~\ref{fig:ml_shap}). The SHAP (Shapley Additive exPlanations \cite{SHAP_NIPS2017_7062}) value of a feature (token) is the average marginal contribution of that feature to the model output. The training set analysis shows which features the model learned to prioritize based on the data it saw. The holdout set analysis shows whether those same priorities hold for new, unseen data.  The features are sorted in descending order by importance. For both configurations \textbf{in} and \textbf{out}, we observe that the top-most essential features in the training step are identical to those in the holdout dataset. For configuration \textbf{in}, this is valid for top-10 features, and for configuration \textbf{out}, the top-12 features (limited to 12 out of a maximum of 66 for printing limitation only). Another valuable observation is that the distributions of importance across the top features are very similar in both the training/holdout datasets, and across both \textbf{in} and \textbf{out} configurations. The importance is distributed evenly across features and is not concentrated in a few. Another observation is that both categorical, numerical, and engineered features, across a wide range, such as electronic, energetic, and physicochemical properties, contribute in accordance with the model targets. These results show a robust, accurate, and reliable model.

Our analyses demonstrate that the fine-tuned transformer is more than a black box, since it internalizes chemically interpretable rules. The most influential tokens reliably represent features such as size mismatch, $d$-electron count, coordination environment, and $d$-band-related descriptors. Essentially, such factors closely resemble the heuristics chemists use to explain trends in core--shell stability and reactivity. In essence, the model systematically evaluates the same competing factors, strain versus coordination, and electronic filling versus hybridization, that underlie the DFT trends in Figs.~\ref{fig2}--\ref{fig4}. This indicates that the model not only extrapolates beyond the computed dataset, but does so in a manner analogous to expert chemical reasoning.

\section{Methods}
	
\subsection{DFT Calculations and Computational Details}
	
All DFT results reported here are obtained from spin-polarized scalar-relativistic calculations\cite{Hohenberg_B864_1964, Kohn_A1133_1965} within the generalized gradient approximation for the exchange-correlation functional according to Perdew--Burke--Ernzerhof\cite{Perdew_3865_1996} (PBE) parametrization. Those calculations are performed with the Vienna \textit{Ab-initio} Simulation Package (VASP)\cite{Kresse_13115_1993, Kresse_11169_1996, Hafner_2044_2008}. The Kohn--Sham (KS) equations are solved with the all-electron projector augmented wave method\cite{Blochl_17953_1994, Kresse_1758_1999}, implemented in VASP\cite{Kresse_13115_1993, Kresse_11169_1996, Hafner_2044_2008}, with KS orbitals expanded in plane-waves up to a cutoff energy of \SI{500}{\electronvolt}, corresponding to a \SI{20}{\percent} larger than the VASP recommended value (ENMAX).

Our nanocluster calculations are simulated in a cubic supercell of size \SI{20}{\angstrom}, which avoids interactions with periodic images, since the distance between them is at least \SI{14}{\angstrom}. On the other hand, the respective free-atoms are simulated in an orthorhombic supercell of size \SI{19}{\angstrom}$\times$\SI{19.25}{\angstrom}$\times$\SI{19.5}{\angstrom}. As nanoclusters and free-atoms, there is no electronic dispersion; the Brillouin zone (BZ) integration was performed with a single $\textbf{k}$-point ($\Gamma$-point). The nanocluster equilibrium geometries were obtained until atomic forces on every atom were smaller than \SI{0.015}{{\electronvolt/}{\angstrom}} and a total energy convergence of \SI{1.0d-6}{\electronvolt} was achieved. The main convergence tests for binary nanoclusters are presented in \SM~(Tables S1$-$S4). At the same time, specific reference results for TM bulk systems (cohesive energy, $E_{\text{coh}}$, and atomic radius, $R_{\text{TM}}$), are calculated and presented in \SM~Fig. S1, in agreement with our previous work\cite{Piotrowski_13172_2024}.
	
\subsection{Atomic Configurations} 

The $xyz$ configurations for \num{13}-atom unary nanoclusters of Group \num{4} (\ce{Ti}, \ce{Zr}, and \ce{Hf}), which consist in the main structural set investigated, and \ce{Fe}, complementary set used for a validation case, are obtained (and re-optimized) as the lowest energy structures from our previous works\cite{Piotrowski_155446_2010, Chaves_15484_2017}. For several \ce{TM13} nanoclusters, the ICO geometry is the most stable configuration, as expected; this is also the case for \ce{Ti13}, \ce{Zr13}, \ce{Hf13}, and \ce{Fe13}\cite{Piotrowski_155446_2010, Chaves_15484_2017}. This quasi-spherical geometry has a central atom and \num{12} (external) equivalent, equidistant atoms, with high-symmetry I$_h$ and a closed-packed structure.

For the binary nanoclusters, we replaced one atom central or external from \ce{X13}, where \ce{X} = \ce{Ti}, \ce{Zr}, \ce{Hf}, or \ce{Fe} by TM (\num{30} elements from the Periodic Table). For the central atom, there is only one option to consider, while for the external case, any of the \num{12} outer atoms can be chosen for replacement, since they are all symmetrically equivalent. Thus, we have two possibilities for each system, with TM preferring to stay inside the \num{12}-atom shell (\ce{X12}), called $in$ (\ce{X$_{12}$TM$^{in}$}), or TM outside, called $out$ (\ce{X$_{12}$TM$^{out}$}), as shown in Fig.~\ref{fig1}. We have performed DFT-PBE calculations to test these two configurational sets: $in$ and $out$, where the $in$ configurations are also treated as core-shells. All unary and binary nanoclusters are relaxed without any symmetry or spin constraint to allow symmetry breaking and a complete potential energy surface exploration. The complete optimized set of structures for \ce{X$_{12}$TM$^{in}$}, \ce{X$_{12}$TM$^{out}$}, and the TM bulk properties are presented in the \SM~(Figs. S1–S9).
	
To characterize the \ce{X$_{12}$TM$^{in}$} and \ce{X$_{12}$TM$^{out}$} systems, we have studied the main energetic, structural, and electronic properties. For example, for elucidated the stability of the systems, in addition to the calculation of vibrational frequencies ($\num{3}N-\num{6}$ vibrational modes, where $N = \num{13}$) as an indication of whether the systems formed are in fact local minima, we have also calculated the binding energy per atom ($E_b$), which is giving by:
\begin{equation} \label{eqn:AE}
\begin{split}
E_b = (E_\text{tot}^{\text{X}_{12}\text{TM}} - 12E_\text{tot}^{\text{X~f-a}} - E_\text{tot}^{\text{TM~f-a}})/13 = \\ 
= (12E_b^u + E_{int} + 12{\Delta}E_{dist})/13~,
\end{split}
\end{equation}
where the first expression is composed by the total energies of the $in$ or $out$ \ce{X$_{12}$TM} systems ($E_\text{tot}^{\text{X}_{12}\text{TM}}$) and the \ce{X} ($E_\text{tot}^{\text{X~f-a}}$) and TM free-atoms ($E_\text{tot}^{\text{TM~f-a}}$), while the second expression is a decomposition of $E_b$ into the binding energy of \ce{X$_{12}$} ICO-derived structure ($E_b^u$), the interaction energy between \ce{X$_{12}$} and TM in the equilibrium geometry of \ce{X$_{12}$TM$^{in}$} or \ce{X$_{12}$TM$^{out}$} ($E_{int}$), and the distortion energy occasioned by interaction between TM species and \ce{X$_{12}$} structure (${\Delta}E_{dist}$). More details about the $E_b$ equation are given in the \SM~ and previous works\cite{Yonezawa_4805_2021, Felix_1040_2023, Piotrowski_13172_2024}.

By subtracting the binding energies of the respective $in$ and $out$ configurations of \ce{X$_{12}$TM}, we can have a direct idea of the TM preference to stay inside the \ce{X$_{12}$} cage, forming the core-shell ($in$), or to stay outside, forming the $out$ configurations. Subtracting the binding energies is equivalent to subtracting the total energies of the configurations of interest, according to the following $\Delta E_{tot}$ equation:
\begin{equation}
\Delta E_{tot} = E_{\text{tot}}^{in} - E_{\text{tot}}^{out}~,
\end{equation}
where $E_{\text{tot}}^{in}$ and $E_{\text{tot}}^{out}$ are the total energies of the \ce{X$_{12}$TM$^{in}$} and \ce{X$_{12}$TM$^{out}$} configurations, respectively. This expression indicates which configuration is more likely, giving preference to the formation of $in$ (core-shell) configurations for negative relative energies and to the formation of $out$ configurations for positive relative energies. Another energy check regarding the stability of binary systems $in$ or $out$ is the formation energy, $E_{form}$, obtained by the equation:
\begin{equation}
E_{form} = E_\text{tot}^{\text{X}_{12}\text{TM}} - (E_\text{tot}^{\text{X}_{12}} + E_\text{tot}^{\text{TM~f-a}})~,
\end{equation}
where $E_\text{tot}^{\text{X}_{12}}$ is the total energy of the lowest energy \ce{X$_{12}$} systems. More details about the energy decomposition equations are presented in \SM. 

Structurally, we used the effective coordination number concept\cite{Hoppe_25_1970, Hoppe_23_1979} to calculate the average bond lengths ($d_{\text{av}}$) and the effective coordination numbers (ECN). For more details, see the \SM. This tool employs a self-consistent approach, utilizing an exponential decay function to derive the effective geometric parameters. Thus, it enables providing an adequate description of the $d_{\text{av}}$ and ECN parameters of a nanocluster, accounting for possible distortions\cite{Piotrowski_155446_2010, Chaves_15484_2017}. 

Electronically, we characterized the binary $in$ and $out$ nanoclusters by obtaining the highest occupied molecular orbital $-$ lowest unoccupied molecular orbital gap (HOMO-LUMO gap) and the center of gravity of the occupied $d$-states ($\epsilon_d$). While the first one is well-defined and consist of the energy difference between the highest energy electronic state that is occupied (HOMO) and the lowest energy electronic state that is unoccupied (LUMO) in a finite system (nanocluster), the second one is based on the $d$-band model proposed by Hammer and N{\o}rskov\cite{Hammer_71_2000}, and can be correlated with the adsorption strength of an adsorbate on the surface of the nanocluster.
	
\subsection{Prediction Model}
	 
The prediction of $E_{form}$ and structural preferences ($in$/$out$ configurations) was conducted through a feature-tokenizer transformer for tabular data (FTTransformer)\cite{TF_gorishniy2021revisiting}. This architecture is designed to handle mixed-type tabular inputs, mapping each categorical and continuous feature to a token in a shared embedding space. These tokens are then processed by stacked self-attention blocks and a multilayer perceptron (MLP) prediction head. We use \num{2968} entries of unary data (QCD) to learn generalizable atomic and feature representations for pretraining, and \num{120} entries of a binary dataset for finetuning and implementing the transfer learning strategy.

\paragraph{Categorical and numerical feature spaces:} Our model uses two main types of features: ($i$) categorical variables, such as atomic numbers for the two sites (TM1, TM2) and discrete periodic-table descriptors for each site and ($ii$) continuous variables, including base structural/energetic terms and engineered physicochemical descriptors. To enrich the input, we augmented the base columns with per-site periodic-table numeric properties, electron-configuration descriptors, and pairwise features such as absolute differences and averages (e.g., $|\Delta \text{.}|$, $\operatorname{avg}(\text{.})$), as well as binary flags for when categorical properties are the same. Categorical features are consistently integer-encoded across both the monometal and bimetal datasets, ensuring fixed embedding sizes. Specifically, we use \num{10} categorical inputs: two atomic-number columns and, for each of the four categorical properties (OxidationStates, StandardState, GroupBlock, ElectronConfiguration), one column per site. Numerical features comprise \num{56} inputs: four base continuous variables (e.g., $n_{\text{atoms},1}$, $n_{\text{atoms},2}$, $coe_1$, $coe_2$); for each of the eight numeric periodic properties (AtomicMass, AtomicRadius, Electronegativity, IonizationEnergy, ElectronAffinity, MeltingPoint, BoilingPoint, Density), we include the pairwise difference and average; for each categorical property, a same/different binary flag; and for the eight electron-configuration descriptors (core\_n\_electron, $3d, 4s, 4d, 4f, 5s, 5d, 6s$), site-wise values for both sites and their pairwise difference and average. In total, this yields \num{66} input features (\num{10} categorical and \num{56} continuous). All continuous inputs and targets are robustly scaled to reduce the influence of outliers and improve optimization stability.

\paragraph{Model architecture and training:} We performed automated hyperparameter optimization with Optuna\cite{optuna_2019}, exploring a range of architectural and training parameters (embedding dimension, number of attention heads, transformer blocks, feed-forward width, dropout rate, activation functions, learning rate, batch size, weight decay, and loss function). Trials were run in parallel using Dask \cite{DASK_matthew_rocklin-proc-scipy-2015}. The top model architectures were manually evaluated for best finetuning performance, and the final decision was made accordingly.

Categorical features are embedded and stacked into a token sequence. Continuous features are normalized and also projected into an embedding dimension, forming a token. The complete token sequence is processed by several transformer blocks (multi-head self-attention and feed-forward layers with dropout and normalization). The output is flattened and passed to an MLP head with dropout, producing two values per sample: $E_{form}^{in}$ and $E_{form}^{out}$. Training uses the Adam optimizer \cite{Kingma_arxiv_2014} with early stopping. Transfer learning is performed by first training the whole model, then freezing the transformer backbone, and fine-tuning using Ridge regression between the transformed representation (embeddings) and the fine-tuning targets.

To quantify uncertainty, we generated calibrated prediction intervals using \num{5}-fold cross-validation (CV+) on the bimetal fine-tuning pool, with the transformer backbone frozen. For each fold, we collect validation residuals and compute holdout predictions, using the median across folds as the final prediction. Intervals are constructed using a fast CV+ mode with per-output residual scoring and a stratification scheme (``mixed Mondrian'') \cite{CP_timans2024adaptive} that groups samples by predicted value or outlier score. Within each group, we compute empirical quantities to set interval widths, ensuring coverage of prediction intervals.

\paragraph{Implementation:} Implementation was structured in a modular configuration in a standalone repository (\url{https://github.com/KIT-Workflows/Nanocluster_Transformers}). The repository holds: ($i$) Jupyter Notebooks for model administration (\texttt{$model_manager.ipynb$}), end-to-end management (\texttt{$manager.ipynb$}), hyperparameter tuning using Optuna (\texttt{$optimization.ipynb$}), and visualization (\texttt{$view.ipynb$}). ($ii$) Python scripts for top-level model declarations (\texttt{$models.py$}), training procedures (\texttt{$ training.py$}), and data preprocessing pipelines. ($iii$) Raw (\texttt{$src/data_actual$}) and processed data directories (\texttt{$src/data_processed$}), final models (\texttt{$src/models_save$}), and temporary models generated during optimization experiments (\texttt{$src/tmp/models$}). The codebase also includes dynamic visualization libraries and a Flask-based graphical user interface (GUI) for model interpretability and user interaction. Overall, this integrated DFT--ML framework provides a predictive, transferable tool for exploring the configurational landscape of bimetallic nanoclusters, enabling rapid screening and design guidance beyond the computed dataset.

\section*{Funding}

The project (HGF ZT-I-PF-5-261 GENIUS) underlying this publication is/was funded by the Initiative and Networking Fund of the Helmholtz Association in the framework of the Helmholtz AI project call.

\section*{Acknowledgements}
This project was in part supported by DFG project WE 1863/41-1 in SPP2363. Authors thank the Rio Grande do Sul Research Foundation (FAPERGS, grant $24/2551-0001551-5$), the Federal District Research Support Foundation (FAPDF, grants $00193-00001817/2023-43$ and $00193-00002073/2023-84$), the National Council for Scientific and Technological Development $-$ CNPq ($303206/2025-0$, $408144/2022-0$, $305174/2023-1$, $444431/2024-1$, $141176/2024-5$, and $444069/2024-0$), and the Coordination for Improvement of Higher Level Education $-$ CAPES (finance Code $001$) for the financial support. Part of this work was performed on the HoreKa supercomputer funded by the Ministry of Science, Research, and the Arts Baden-Württemberg and by the Federal Ministry of Education and Research. A.C.D. also acknowledges funding from PDPG-FAPDF-CAPES Centro-Oeste grant number $00193-00000867/2024-94$.

\section*{Data Availability Statement}
	
The data that supports the findings of this study are available at \url{https://github.com/KIT-Workflows/Nanocluster_Transformers} or from the corresponding author upon reasonable request.
	
\section*{Code Availability}
	
The code supporting this study's findings is available at \url{https://github.com/KIT-Workflows/Nanocluster_Transformers}.

\bibliographystyle{unsrtnat}
\bibliography{jshort, boxref}

@article{Bastos_2025,
  title = {First-principles statistical investigation of thermodynamic behavior with excitonic effects in Mo1−xWxSe2  alloys through a data-driven workflow approach},
  ISSN = {2050-7496},
  url = {http://dx.doi.org/10.1039/D5TA02721G},
  DOI = {10.1039/d5ta02721g},
  journal = {Journal of Materials Chemistry A},
  publisher = {Royal Society of Chemistry (RSC)},
  author = {Bastos,  Carlos Maciel de O. and C. Dias,  Alexandre and Ferreira de Brito,  Ana Carolina and Barcelos,  Ingrid D. and da Rosa,  Andréia Luisa and Silveira,  Danilo Neves and Piotrowski,  Maurício J. and Wenzel,  Wolfgang and R\^ego,  Celso R. C. and Guedes-Sobrinho,  Diego},
  year = {2025}
}

@Article{Soleymanibrojeni_2249_2024,
  author    = {Soleymanibrojeni, Mohammad and Caldeira Rego, Celso Ricardo and Esmaeilpour, Meysam and Wenzel, Wolfgang},
  title     = {An active learning approach to model solid-electrolyte interphase formation in {Li}-ion batteries},
  journal   = JMaterChemA,
  year      = {2024},
  volume    = {12},
  number    = {4},
  pages     = {2249–2266},
  publisher = {Royal Society of Chemistry (RSC)},
}

@Article{Dalmedico_e202400118_2024,
  author    = {Dalmedico, J. F. and Silveira, D. N. and O. de Araujo, L. and Wenzel, W. and R\^ego, C. R. C. and Dias, A. C. and Guedes‐Sobrinho, D. and Piotrowski, Maurício J.},
  title     = {Tuning Electronic and Structural Properties of Lead‐Free Metal Halide Perovskites: A Comparative Study of 2{D} {R}uddlesden‐{P}opper and 3{D} Compositions},
  journal   = ChemPhysChem,
  year      = {2024},
  volume    = {25},
  number    = {16},
  pages     = {e202400118},
  publisher = {Wiley},
}

@Article{Schaarschmidt_2102638_2021,
  author    = {Schaarschmidt, Joerg and Yuan, Jie and Strunk, Timo and Kondov, Ivan and Huber, Sebastiaan P. and Pizzi, Giovanni and Kahle, Leonid and B\"{o}lle, Felix T. and Castelli, Ivano E. and Vegge, Tejs and Hanke, Felix and Hickel, Tilmann and Neugebauer, J\"{o}rg and R\^ego, Celso R. C. and Wenzel, Wolfgang},
  title     = {Workflow Engineering in Materials Design within the BATTERY 2030+ Project},
  journal   = AdvEnergyMater,
  year      = {2021},
  volume    = {12},
  number    = {17},
  pages     = {2102638},
  publisher = {Wiley},
}

@Article{Bekemeier_XXXX_2025,
  author    = {Bekemeier, Simon and Caldeira R\^ego, Celso Ricardo and Mai, Han Lin and Saikia, Ujjal and Waseda, Osamu and Apel, Markus and Arendt, Felix and Aschemann, Alexander and Bayerlein, Bernd and Courant, Robert and Dziwis, Gordian and Fuchs, Florian and Giese, Ulrich and Junghanns, Kurt and Kamal, Mohamed and Koschmieder, Lukas and Leineweber, Sebastian and Luger, Marc and Lukas, Marco and Maas, J\"{u}rgen and Mertens, Jana and Mieller, Bj\"{o}rn and Overmeyer, Ludger and Pirch, Norbert and Reimann, Jan and Schr\"{o}ck, Sebastian and Schulze, Philipp and Schuster, J\"{o}rg and Seidel, Alexander and Shchyglo, Oleg and Sierka, Marek and Silze, Frank and Stier, Simon and Tegeler, Marvin and Unger, J\"{o}rg F. and Weber, Matthias and Hickel, Tilmann and Schaarschmidt, J\"{o}rg},
  title     = {Advancing Digital Transformation in Material Science: The Role of Workflows Within the MaterialDigital Initiative},
  journal   = AdvEngMater,
  year      = {2025},
  publisher = {Wiley},
}

@Article{Rego_2022,
  author    = {Celso R. C. R{\^{e}}go and Jörg Schaarschmidt and Tobias Schlöder and Montserrat Penaloza-Amion and Saientan Bag and Tobias Neumann and Timo Strunk and Wolfgang Wenzel},
  title     = {{SimStack}: An Intuitive Workflow Framework},
  journal   = FrontMater,
  year      = {2022},
  volume    = {9},
  publisher = {Frontiers Media {SA}},
}

@Article{Rego_XXXX_2025,
  author    = {R\^ego, Celso R. C. and Wenzel, Wolfgang and Piotrowski, Maurício J. and Dias, Alexandre C. and Maciel de Oliveira Bastos, Carlos and de Araujo, Luis O. and Guedes-Sobrinho, Diego},
  title     = {Digital workflow optimization of van der {W}aals methods for improved halide perovskite solar materials},
  journal   = DigitDiscov,
  year      = {2025},
  publisher = {Royal Society of Chemistry (RSC)},
}

@article{optuna_2019,
title={Optuna: A Next-generation Hyperparameter Optimization Framework},
author={Akiba, Takuya and Sano, Shotaro and Yanase, Toshihiko and Ohta, Takeru and Koyama, Masanori},
journal={Proceedings of the 25th {ACM} {SIGKDD} International Conference on Knowledge Discovery and Data Mining},
pages={2623--2631},
year={2019},
doi={10.1145/3292500.3330701}
}

@article{SHAP_NIPS2017_7062,
title = {A Unified Approach to Interpreting Model Predictions},
author = {Lundberg, Scott M and Lee, Su-In},
journal = {Advances in Neural Information Processing Systems},
volume = {30},
pages = {4765--4774},
year = {2017}
}

@article{DASK_matthew_rocklin-proc-scipy-2015,
author    = { Matthew Rocklin },
title     = { Dask: Parallel Computation with Blocked algorithms and Task Scheduling },
booktitle = { Proceedings of the 14th Python in Science Conference },
pages     = { 130 - 136 },
year      = { 2015 },
editor    = { Kathryn Huff and James Bergstra }
}

@article{CP_timans2024adaptive,
  title={Adaptive bounding box uncertainties via two-step conformal prediction},
  author={Timans, Alexander and Straehle, Christoph-Nikolas and Sakmann, Kaspar and Nalisnick, Eric},
  booktitle={European Conference on Computer Vision},
  pages={363--398},
  year={2024},
  organization={Springer}
}

@article{TF_gorishniy2021revisiting,
  title={Revisiting deep learning models for tabular data},
  author={Gorishniy, Yury and Rubachev, Ivan and Khrulkov, Valentin and Babenko, Artem},
  journal={Advances in neural information processing systems},
  volume={34},
  pages={18932--18943},
  year={2021}
}

@Article{Manna_308_2023,
author  = {S. Manna and Y. Wang and A. Hernandez and P. Lile and S. Liu and T. Mueller},
journal = ScientData,
title   = {A database of low-energy atomically precise nanoclusters},
year    = {2023},
pages   = {308},
volume  = {10},
doi     = {10.1038/s41597-023-02200-4}
}

@book{Harrison_2012,
title = {Electronic structure and the properties of solids: the physics of the chemical bond},
author = {W. A. Harrison},
year = {2012},
Publisher = {Courier Corporation}
}

@Article{Kaya_37209_2023,
author  = {D. Kaya and I. Demiroglu and I. B. Isik and H. H. Isik and S. K. \c{C}etin and C. Sevik and A. Ekicibil and F. Karadag},
journal = IntJHydrogenEnergy,
title   = {Highly active bimetallic Pt-Cu nanoparticles for the electrocatalysis of hydrogen evolution reactions: Experimental and theoretical insight},
year    = {2023},
pages   = {37209--37223},
volume  = {48},
doi     = {10.1016/j.ijhydene.2023.06.100}
}

@Article{Wu_78_2019,
author  = {C. H. Wu and C. Liu and D. Su and H. L. Xin and H.-T. Fang and B. Eren and S. Zhang and C. B. Murray and M. B. Salmeron},
journal = NatCatal,
title   = {Bimetallic synergy in cobalt-palladium nanocatalysts for CO oxidation},
year    = {2019},
pages   = {78--85},
volume  = {2},
doi     = {10.1038/s41929-018-0190-6}
}

@Article{Chakraborty_8208_2017,
author  = {I. Chakraborty and T. Pradeep},
journal = ChemRev,
title   = {Atomically precise clusters of noble metals: Emerging link between atoms and nanoparticles},
year    = {2017},
pages   = {8208--8271},
volume  = {117},
doi     = {10.1021/acs.chemrev.6b00769}
}

@Article{Ke_806_2025,
author  = {S. Ke and Y. Zhao and X. Min and X. Zhu and X. Li and B. Yang and F. Yang and X. Wu and R. Mi and Y. Liu Z. Huang and M. Fang},
journal = IntJHydrogenEnergy,
title   = {Tailoring the d-band center in Pt-based catalysts for hydrogen evolution via transition metals incorporation},
year    = {2025},
pages   = {806--816},
volume  = {105},
doi     = {10.1016/j.ijhydene.2025.01.303}
}

@Article{Yamamoto_116801_2006,
author  = {Y. Yamamoto and T. Miura and M. Suzuki and N. Kawamura and H. Miyagawa and T. Nakamura and K. Kobayashi and T. Teranishi and H. Hori},
journal = PhysRevLett,
title   = {Direct Observation of Ferromagnetic Spin Polarization in Gold Nanoparticles},
year    = {2006},
pages   = {116801},
volume  = {93},
doi     = {10.1103/PhysRevLett.93.116801}
}

@Article{Pereiro_063204_2007,
author  = {M. Pereiro and D. Baldomir and J. E. Arias},
journal = PhysRevA,
title   = {Unexpected magnetism of small silver clusters},
year    = {2007},
pages   = {063204},
volume  = {75},
doi     = {10.1103/PhysRevA.75.063204}
}

@Article{Gawande_7540_2015,
author  = {M. B. Gawande and A. Goswami and T. Asefa and H. Guo and A. V. Biradar and D.-L. Peng and R. Zboril and R. S. Varma},
journal = ChemSocRev,
title   = {Core-shell nanoparticles: synthesis and applications in catalysis and electrocatalysis},
year    = {2015},
pages   = {7540--7590},
volume  = {44},
doi    = {10.1039/C5CS00343A},
}

@Article{Huang_arxiv_2020,
title={Tabtransformer: Tabular data modeling using contextual embeddings},
author={Huang, Xin and Khetan, Ashish and Cvitkovic, Milan and Karnin, Zohar},
journal={arXiv preprint arXiv:2012.06678},
year={2020}
}

@Article{Kingma_arxiv_2014,
title={Adam: A method for stochastic optimization},
author={Kingma, Diederik P and Ba, Jimmy},
journal={arXiv preprint arXiv:1412.6980},
year={2014}
}

@Book{Hammer_71_2000,
author = {B. Hammer and J. K. N{\o}rskov},
title = {Advances in Catalysis},
publisher = {Academic Press Inc, San Diego},
year = {2000},
doi = { }
}

@Article{Hammer_211_1995,
author  = {B. Hammer and J. K. N{\o}rskov},
journal = SurfSci,
title   = {Electronic Factors Determining the Reactivity of Metal Surfaces},
year    = {1995},
pages   = {211--220},
volume  = {343},
doi    = {10.1016/0039-6028(96)80007-0},
}

@Article{Piotrowski_13172_2024,
author = {M. J. Piotrowski and J. M. T. Palheta and R. Fournier},
title = {Cage doping of Ti, Zr, and Hf-based 13-atom nanoclusters: two sides of the same coin},
journal = PhysChemChemPhys,
volume = {26},
pages = {13172--13181},
year = {2024},
doi = {10.1039/d4cp00518j}
}

@Article{Alonso_637_2000,
author = {Alonso, J. A.},
title = {{E}lectronic and {A}tomic {S}tructure, and {M}agnetism of {T}ransition-{M}etal {C}lusters},
journal = ChemRev,
volume = {100},
pages = {637--678},
year = {2000},
doi = {10.1021/cr980391o}
}

@Article{Aiken_1_1999,
Title = {A review of modern transition-metal nanoclusters: their synthesis, characterization, and applications in catalysis},
author = {J. D. Aiken III and R. G. Finke},
journal = JMolCatalAChem,
volume = {145},
pages = {1--44},
year = {1999},
doi = {10.1016/S1381-1169(99)00098-9}
}

@Article{Mackay_916_1962,
title = {{A} {D}ense {N}on-{C}rystallographic {P}acking of {E}qual {S}pheres},
author = {A. L. Mackay},
doi = {10.1107/s0365110x6200239x},
journal = ActaCrystallogr, 
pages = {916--918},
volume = {15},
year = {1962}
}

@Article{Baletto_371_2005,
author = {Francesca Baletto and Riccardo Ferrando},
title = {{S}tructural {P}roperties of {N}anoclusters: {E}nergetic, {T}hermodynamic, and {K}inetic {E}ffects},
journal = RevModPhys,
volume = {77},
pages = {371--423},
year = {2005},
doi = {10.1103/revmodphys.77.371}
}

@Article{Ferrando_845_2008,
title = {{N}anoalloys: {F}rom {T}heory to {A}pplications of {A}lloy {C}lusters and {N}anoparticles},
author = {R. Ferrando and J. Jellinek and R. L. Johnston},
journal = ChemRev,
volume = {108},
pages = {845--910},
year = {2008},
doi = {10.1021/cr040090g}
}

@Article{Toshima_1179_1998,
author  = {N. Toshima and T. Yonezawa},
journal = NewJChem,
title   = {Bimetallic Nanoparticles - Novel Materials for Chemical and Physical Applications},
year    = {1998},
pages   = {1179--1201},
volume  = {22},
doi     = {10.1039/A805753B}
}

@Article{Eom_8883_2021,
title = {General trends in core--shell preferences for bimetallic nanoparticles},
author = {N. Eom and M. E. Messing and J. Johansson and K. Deppert},
journal = ACSNano,
volume = {15},
pages = {8883--8895},
year = {2021},
doi = {10.1021/acsnano.1c01500}
}

@Article{Batista_7431_2019,
title = {Adsorption of CO, NO, and H$_2$ on the Pd$_n$Au$_{55-n}$ Nanoclusters: A Density Functional Theory Investigation within the van der Waals D3 Corrections},
author = {K. E. A. Batista and J. L. F. {Da Silva} and M. J. Piotrowski},
journal = JPhysChemC,
volume = {123},
pages = {7431--7439},
year = {2019},
doi = {10.1021/acs.jpcc.8b12219}
}

@Article{Blochl_17953_1994,
author = {P. E. Bl{\"{o}}chl},
title = {Projector Augmented-Wave Method},
journal = PhysRevB,
volume = {50},
pages = {17953--17979},
year = {1994},
doi = {10.1103/PhysRevB.50.17953}
}

@Article{Castleman_2664_2009, 
author = {A. W. {Castleman,~Jr.} and S. N. Khanna},
title = {Clusters, Superatoms, and Building Blocks of New Materials},
journal = JPhysChemC,
volume = {113},
pages = {2664--2675},
year = {2009},
doi = {10.1021/jp806850h}
}

@article{Chaves_15484_2017,
author = {A. S. Chaves and M. J. Piotrowski and J. L. F. Da Silva},
title = {Evolution of the Structural, Energetic, and Electronic Properties of the 3$d$, 4$d$, and 5$d$ Transition-Metal Clusters (30 {TM}$_{n}$ Systems for $n$= 2--15): A Density Functional Theory Investigation},
journal = PhysChemChemPhys,
volume = {19},
number = {23},
pages = {15484--15502},
year = {2017},
doi = {10.1039/C7CP02240A}
}

@Article{Felix_1040_2023,
author = {J. P. C. S. Felix and K. E. A. Batista and W. O. Morais and G. R. Nagurniak and R. P. Orenha and C. R. C. R{\^e}go and D. Guedes-Sobrinho and R. L. T. Parreira and M. M. Ferrer and M. J. Piotrowski},
title = {Molecular adsorption on coinage metal subnanoclusters: A {DFT}+{D}3 investigation},
journal = JComputChem,
volume = {44},
pages = {1040--1051},
year = {2023},
doi = {10.1002/jcc.27063}
}

@Article{Fernando_6112_2015,
author = {A. Fernando and K. L. Dimuthu and M. Weerawardene and N. V. Karimova and C. M. Aikens},
title = {Quantum Mechanical Studies of Large Metal, Metal Oxide, and Metal Chalcogenide Nanoparticles and Clusters},
journal = ChemRev,
volume = {115},
pages = {6112--6216},
year = {2015},
doi = {10.1021/cr500506r}
}

@Article{Hafner_2044_2008,
author = {J. Hafner},
title = {${A}b-initio$ simulations of materials using {VASP}: Density-functional theory and beyond},
journal = JComputChem,
volume = {29},
pages = {2044--2078},
year = {2008},
doi = {10.1002/jcc.21057}
}

@Article{Hohenberg_B864_1964,
author = {P. Hohenberg and W. Kohn},
title = {Inhomogeneous Electron Gas},
journal = PhysRev,
volume = {136},
pages = {B864--B871},
year = {1964},
doi = {10.1103/PhysRev.136.B864}
}

@Article{Hoppe_25_1970,
author = {R. Hoppe},
title = {The Coordination Number $-$ an ''Inorganic Chameleon''},
journal = AngewChemIntEd,
volume = {9},
pages = {25--34},
year = {1970},
doi = {10.1002/anie.197000251}
}

@Article{Hoppe_23_1979,
title = {{E}ffective {C}oordination {N}umbers ({ECoN}) and {M}ean {A}ctive {F}ictive {I}onic {R}adii ({MEFIR})},
author = {R. Hoppe},
journal = ZKristallogr, 
volume = {150},
pages = {23--52},
year = {1979},
doi = {10.1524/zkri.1979.150.1-4.23}
}

@Article{Kohn_A1133_1965,
author = {W. Kohn and L. J. Sham},
title = {Self-Consistent Equations Including Exchange and Correlation Effects},
journal = PhysRev,
volume = {140},
pages = {A1133--A1138},
year = {1965},
doi = {10.1103/PhysRev.140.A1133}
}

@Article{Kresse_13115_1993,
author = {G. Kresse and J. Hafner},
title = {Ab Initio Molecular Dynamics for Open-Shell Transition Metals},
journal = PhysRevB,
volume = {48},
pages = {13115--13126},
year = {1993},
doi = {10.1103/PhysRevB.48.13115}
}

@Article{Kresse_11169_1996,
author = {G. Kresse and J. Furthm{\"{u}}ller},
title = {Efficient Iterative Schemes for Ab Initio Total-Energy Calculations Using a Plane-Wave Basis Set},
journal = PhysRevB,
volume = {54},
pages = {11169--11186},
year = {1996},
doi = {10.1103/PhysRevB.54.11169}
}

@Article{Kresse_1758_1999,
author = {G. Kresse and D. Joubert},
title = {From Ultrasoft Pseudopotentials to the Projector Agumented-Wave Method}, 
journal = PhysRevB,
volume = {59},
pages = {1758--1775},
year = {1999},
doi = {10.1103/PhysRevB.59.1758}
}

@Article{Mendes_1158_2020,
author = {P. C. D. Mendes and S. G. Justo and J. Mucelini and M. D. Soares and K. E. A. Batista and M. G. Quiles and M. J. Piotrowski and J. L. F. Da Silva},
title = {Ab Initio Insights into the Formation Mechanisms of 55-Atom Pt-Based Core-Shell Nanoalloys},
journal = JPhysChemC,
volume = {124},
pages = {1158--1164},
year = {2020},
doi = {10.1021/acs.jpcc.9b09561}
}

@Article{Perdew_3865_1996,
author = {J. P. Perdew and K. Burke and M. Ernzerhof},
title = {Generalized Gradient Approximation Made Simple},
journal = PhysRevLett,
volume = {77},
pages = {3865--3868},
year = {1996},
doi = {10.1103/PhysRevLett.77.3865}
}

@Article{Piotrowski_155446_2010,
author = {M. J. Piotrowski and P. Piquini and J. L. F. Da Silva}, 
title = {Density Functional Theory Investigation of $3d$, $4d$, and $5d$ 13-atom Metal Clusters},
journal = PhysRevB,
volume = {81},
pages = {155446},
year = {2010},
doi = {10.1103/PhysRevB.81.155446}
}

@Article{Wang_14023_2009,
author = {L.-L. Wang and D. D. Johnson},
title = {Predict Trends of Core-Shell Preferences for 132 Late Transition-Metal Binary-Alloy Nanoparticles},
journal = JAmChemSoc,
volume = {131},
pages = {14023--14029},
year = {2009},
doi = {10.1021/ja903247x}
}

@Article{Yonezawa_4805_2021,
author = {A. F. Yonezawa and G. R. Nagurniak and R. P. Orenha and E. H. da Silva and R. L. T. Parreira and M. J. Piotrowski},
title = {Stability Changes in Iridium Nanoclusters via Monoxide Adsorption: A DFT Study within the van der Waals Corrections},
journal = JPhysChemA,
volume = {125},
pages = {4805--4818},
year = {2021},
doi = {10.1021/acs.jpca.1c02694}
}

@string{ACSNano = "ACS Nano"}

@string{ActaCrystallogr = "Acta Crystallogr."}

@string{AdvEnergyMater = "Adv. Energy Mater."}

@string{AngewChemIntEd = "Angew. Chem. Int. Ed."}

@string{ChemPhysChem = "ChemPhysChem"}

@string{ChemRev = "Chem. Rev."}

@string{ChemSocRev = "Chem. Soc. Rev."}

@string{IntJHydrogenEnergy = "Int. J. Hydrogen Energy"}

@string{JAmChemSoc = "J. Am. Chem. Soc."}

@string{jan = "Jan."}

@string{JComputChem = "J. Comput. Chem."}

@string{JMaterChemA = "J. Mater. Chem. A"}

@string{JMolCatalAChem = "J. Mol. Catal. A: Chem."}

@string{JPhysChemA = "J. Phys. Chem. A"}

@string{JPhysChemC = "J. Phys. Chem. C"}

@string{NatCatal = "Nat. Catal."}

@string{NewJChem = "New. J. Chem."}

@string{PhysChemChemPhys = "Phys. Chem. Chem. Phys."}

@string{PhysRevA = "Phys. Rev. A"}

@string{PhysRevB = "Phys. Rev. B"}

@string{PhysRevLett = "Phys. Rev. Lett."}

@string{PhysRev = "Phys. Rev."}

@string{RevModPhys = "Rev. Mod. Phys."}

@string{Science = "Science"}

@string{ScientData = "Scientific Data"}

@string{Small = "Small"}

@string{SurfSci = "Surf. Sci."}

@string{ZKristallogr = "Z. Kristallogr."}

\clearpage
    
\section*{Author information}
    
\subsection*{Contributions}
    
JMTP and MJP performed the DFT calculations, analyzed the data, and prepared the figures. MS, ORF, and CRCR curated the data and developed the PAI algorithm code. CRCR and MS supervised and refined the ML coding for the final version. JMTP, MJP, MS, and CRCR drafted the initial version of the manuscript. ACD, DGS, WW, and RA contributed to data curation and critically reviewed the manuscript findings. MJP and CRCR jointly supervised the project and led manuscript preparation. All authors contributed to the writing and revision of the manuscript.

\section*{Ethics declarations}

\subsection*{Competing Interests}
	
All authors declare no competing financial or non-financial interests. 

\end{document}


\newpage

\maketitle

\newpage

\tableofcontents

\maketitle

\newpage
\clearpage

\section{Convergence Tests}

We performed methodological tests to obtain the main parameters used in our calculations. In Tables S1, S2, S3, and S4 we show some properties (the relative total energy -- $\Delta E_\text{tot} = E_\text{tot}^\text{A} - E_\text{tot}^\text{B}$, average bond length -- $d_{av}^\text{A}$ and $d_{av}^\text{B}$, effective coordination number -- ECN$^\text{A}$ and ECN$^\text{B}$, and total magnetic moment -- $m_\text{tot}^\text{A}$ and $m_\text{tot}^\text{B}$) for the arbitrary cases of the \ce{Hf13} (named A) and \ce{Hf12Hg} (named B) as a function of the parameters.

\begin{table*}[ht!]
\caption*{Table S1: {\sl{Box size}}: the relative total energy ($\Delta E_\text{tot}$), average bond length ($d_{av}^\text{A}$ and $d_{av}^\text{B}$), effective coordination number (ECN$^\text{A}$ and ECN$^\text{B}$), and total magnetic moment ($m_\text{tot}^\text{A}$ and $m_\text{tot}^\text{B}$).
}
\centering
\definecolor{graylight}{gray}{0.8}
\scalebox{0.9}{
\begin{tabular}{c|c|c|c|c|c|c|c}
		\hline 
		\textbf{Box Size (\AA)} & $\mathbf{\Delta E_{tot}}$ \textbf{(eV)} & $\mathbf{ECN^A}$ & $\mathbf{ECN^B}$ & $\mathbf{d_{av}^A}$ \textbf{(\AA)} & $\mathbf{d_{av}^B}$ \textbf{(\AA)} & $\mathbf{m_{tot}^A}$ ($\mathbf{\mu_A}$) & $\mathbf{m_{tot}^B}$ ($\mathbf{\mu_B}$) \\ 
		\hline 
		12 & -12.4168 & 6.3730 & 6.3914 & 2.2093 & 2.9398 & 6.0000 & 2.0000 \\
		14 & -12.4939 & 6.3722 & 6.3941 & 2.9461 & 2.9382 & 6.0000 & 2.0000 \\		
		16 & -12.4236 & 6.3730 & 6.3913 & 2.9457 & 2.9389 & 6.0000 & 2.0000 \\ 		
		18 & -12.4608 & 6.3731 & 6.3945 & 2.9457 & 2.9396 & 6.0000 & 2.0000 \\ 
        \rowcolor{gray!30}
		20 & -12.4599 & 6.3729 & 6.3945 & 2.9450 & 2.9398 & 6.0000 & 2.0000 \\
		22 & -12.4337 & 6.3702 & 6.3946 & 2.9441 & 2.9398 & 6.0000 & 2.0000 \\ 		
		24 & -12.4271 & 6.3724 & 6.3913 & 2.9456 & 2.9389 & 6.0000 & 2.0000 \\ 
		\hline
\end{tabular}
}
\label{t1}
\end{table*}

\begin{table*}[ht!]
\caption*{Table S2: ENCUT = energy-cut-off: X $\cdot$ ENMAX, where the ENMAX = 282.964 eV, the relative total energy ($\Delta E_\text{tot}$), average bond length ($d_{av}^\text{A}$ and $d_{av}^\text{B}$), effective coordination number (ECN$^\text{A}$ and ECN$^\text{B}$), and total magnetic moment ($m_\text{tot}^\text{A}$ and $m_\text{tot}^\text{B}$).
}
\centering
\scalebox{0.85}{
 	\begin{tabular}{c|c|c|c|c|c|c|c}
		\hline 
		\textbf{X $\cdot$ ENMAX} \textbf{(eV)} & $\mathbf{\Delta E_{tot}}$ \textbf{(eV)} & $\mathbf{ECN^A}$ & $\mathbf{ECN^B}$ & $\mathbf{d_{av}^A}$ \textbf{(\AA)} & $\mathbf{d_{av}^B}$ \textbf{(\AA)} & $\mathbf{m_{tot}^A}$ ($\mathbf{\mu_A}$) & $\mathbf{m_{tot}^B}$ ($\mathbf{\mu_B}$) \\ 
		\hline 
		0.500 & -10.4053 & 6.1700 & 6.3569 & 2.8254 & 2.7650 & 6.0000 & 2.0000 \\ 
		0.750 & -12.3137 & 6.3542 & 6.3587 & 2.9876 & 2.9451 & 6.0000 & 2.0000 \\  
		1.000 & -12.3937 & 6.3776 & 6.3901 & 2.9423 & 2.9341 & 6.0000 & 2.0000 \\ 
		1.125 & -12.4290 & 6.3733 & 6.3923 & 2.9461 & 2.9402 & 6.0000 & 2.0000 \\
        \rowcolor{gray!30}
		1.250 & -12.4594 & 6.3731 & 6.3945 & 2.9459 & 2.9398 & 6.0000 & 2.0000 \\  
		1.500 & -12.4279 & 6.3730 & 6.3920 & 2.9456 & 2.9388 & 6.0000 & 2.0000 \\  
		2.000 & -12.4362 & 6.3729 & 6.3913 & 2.9450 & 2.9383 & 6.0000 & 2.0000 \\ 
		\hline 
	\end{tabular}\label{t2}
    }
 \end{table*}

\newpage

\begin{table*}[ht!]
\caption*{Table S3: EDIFF = energy difference: electronic convergence, the relative total energy ($\Delta E_\text{tot}$), average bond length ($d_{av}^\text{A}$ and $d_{av}^\text{B}$), effective coordination number (ECN$^\text{A}$ and ECN$^\text{B}$), and total magnetic moment ($m_\text{tot}^\text{A}$ and $m_\text{tot}^\text{B}$).
}
\centering
\scalebox{0.9}{
\begin{tabular}{cccccccc}
		\hline 
		\textbf{EDIFF (eV)} & $\mathbf{\Delta E_{tot}}$ \textbf{(eV)} & $\mathbf{ECN^A}$ & $\mathbf{ECN^B}$ & $\mathbf{d_{av}^A}$ \textbf{(\AA)} & $\mathbf{d_{av}^B}$ \textbf{(\AA)} & $\mathbf{m_{tot}^A}$ ($\mathbf{\mu_A}$) & $\mathbf{m_{tot}^B}$ ($\mathbf{\mu_B}$) \\ 
		\hline 
		$10^{-2}$ & -12.4600 & 6.3729 & 6.3946 & 2.9454 & 2.9386 & 6.0000 & 2.0000 \\ 
		$10^{-3}$ & -12.4598 & 6.3740 & 6.3943 & 2.9457 & 2.9385 & 6.0000 & 2.0000 \\ 
		$10^{-4}$ & -12.4600 & 6.3726 & 6.3939 & 2.9459 & 2.9394 & 6.0000 & 2.0000 \\ 
		$10^{-5}$ & -12.4600 & 6.3731 & 6.3946 & 2.9457 & 2.9391 & 6.0000 & 2.0000 \\
        \rowcolor{gray!30}
		$10^{-6}$ & -12.4599 & 6.3729 & 6.3945 & 2.9450 & 2.9398 & 6.0000 & 2.0000 \\ 
		$10^{-7}$ & -12.4600 & 6.3731 & 6.3947 & 2.9456 & 2.9398 & 6.0000 & 2.0000 \\ 
		$10^{-8}$ & -12.4599 & 6.3729 & 6.3946 & 2.9450 & 2.9398 & 6.0000 & 2.0000 \\ 
		\hline 
	\end{tabular} \label{t3}
    }
\end{table*}

\begin{table*}[ht!]
\caption*{Table S4: EDIFFG = energy difference gradient: ionic convergence, the relative total energy ($\Delta E_\text{tot}$), average bond length ($d_{av}^\text{A}$ and $d_{av}^\text{B}$), effective coordination number (ECN$^\text{A}$ and ECN$^\text{B}$), and total magnetic moment ($m_\text{tot}^\text{A}$ and $m_\text{tot}^\text{B}$).
}
\label{ediffg}
\centering
\scalebox{0.85}{
\begin{tabular}{cccccccc}
	\hline 
	\textbf{EDIFFG (eV/\AA)} & $\mathbf{\Delta E_{tot}}$ \textbf{(eV)} & $\mathbf{ECN^A}$ & $\mathbf{ECN^B}$ & $\mathbf{d_{av}^A}$ \textbf{(\AA)} & $\mathbf{d_{av}^B}$ \textbf{(\AA)} & $\mathbf{m_{tot}^A}$ ($\mathbf{\mu_A}$) & $\mathbf{m_{tot}^B}$ ($\mathbf{\mu_B}$) \\ 
		\hline 
	-0,100 & -12.4265 & 6.3729 & 6.3918 & 2.9450 & 2.9389 & 6.0000 & 2.0000 \\  
	-0,050 & -12.4265 & 6.3729 & 6.3918 & 2.9450 & 2.9389 & 6.0000 & 2.0000 \\ 
	-0,025 & -12.4265 & 6.3729 & 6.3918 & 2.9450 & 2.9389 & 6.0000 & 2.0000 \\ 
	-0,020 & -12.4265 & 6.3729 & 6.3918 & 2.9450 & 2.9389 & 6.0000 & 2.0000 \\ 
	-0,015 & -12.4266 & 6.3731 & 6.3918 & 2.9457 & 2.9389 & 6.0000 & 2.0000 \\
    \rowcolor{gray!30}
	-0,010 & -12.4266 & 6.3732 & 6.3918 & 2.9461 & 2.9389 & 6.0000 & 2.0000 \\ 
	-0,005 & -12.4265 & 6.3731 & 6.3913 & 2.9460 & 2.9389 & 6.0000 & 2.0000 \\ 
	\hline 
\end{tabular} \label{t4}
} 
\end{table*}

\newpage

\section{Proofs for the Energetic Equations}

Below, we have provided the equation definitions used in our energetic analysis for the \ce{$X_{12}TM^{in}$} and  \ce{$X_{12}TM^{out}$} systems, where $X$ are the Ti, Zr, Hf, and Fe atoms and TM the transition metal atoms, as well as, the proofs that demonstrate the equivalence of the equations for the alternative and usual binding.

\textbf{Binding Energy} $\,\,\,$

The usual equation of the binding energy per atom for all systems is given by,

\begin{equation}
	E_b = \dfrac{E_{tot}^{X_{12}TM} - 12 E_{tot}^{X-free} - E_{tot}^{TM - free}}{13}, \label{Eb}
\end{equation}

where $E_{tot}^{X_{12}TM}$ is the total energy of the systems, and $E_{tot}^{X-free}$ is total energy of the Ti, Zr and Hf free atoms, $E_{tot}^{TM - free}$ is total energy of the TM free atom.

\textbf{Interaction Energy} $\,\,\,$

The equation gives the interaction energy,

\begin{equation}
	E_{int} = E_{tot}^{X_{12}TM} - E_{tot}^{X_{12} - frozen} - E_{tot}^{TM - free}, \label{Eint}
\end{equation}

where $E_{tot}^{X_{12} - frozen}$ is the total energy of the frozen system, \ce{X$_{12}$}, without TM atoms and $E_{tot}^{TM - free}$ is the total energy of the TM free atom.

\textbf{Relative Distortion Energy} $\,\,\,$

The relative distortion energy is the energy differences per atom between the frozen and relaxed systems. The distortion energy between the frozen \ce{E$_{tot}^{X_{12} - frozen}$} system, without TM atom, and the optimized \ce{$E_{tot}^{X_{12}}$} nanocluster is given by,

\begin{equation}
	\Delta E_{dist} = \dfrac{E_{tot}^{X_{12} - frozen} - E_{tot}^{X_{12}}}{12}, \label{Edist}
\end{equation}

\textbf{Alternative Equation for the Binding Energy}

Is it possible to write an alternative equation for the binding energy of the nanocluster systems? It is composed of the binding energies of the unprotected $E_b^u$ nanoclusters, without TM atom, given by,

\begin{equation}
	E_b^u = \dfrac{E_{tot}^{X_{12}} - 12 E_{tot}^{X-livre}}{12}, \label{Eub}
\end{equation}

\noindent with the interaction energy (\ref{Eint}) and the relative distortion energy (\ref{Edist}) of systems, we can write an alternative equation for binding energy as,

\begin{equation}
	E_{b} = \dfrac{12 E^u_b + E_{int} + 12 \Delta E_{dist}}{13}, \label{E_ab}
\end{equation}

To prove the equivalence of both equations, (\ref{Eb}) and (\ref{E_ab}), we replace equations (\ref{Eub}), (\ref{Eint}) and (\ref{Edist}) in equation (\ref{E_ab}), consequently,

\begin{equation}
	E_{b} = \frac{12 \left(\frac{E_{tot}^{X_{12}} - 12 E_{tot}^{X-free}}{12} \right) + E_{tot}^{X_{12}TM} - E_{tot}^{X_{12}-frozen} - E_{tot}^{TM - free} + 12 \left(\frac{E_{tot}^{X_{12} - frozen} - E_{tot}^{X_{12}} }{12}\right)}{13},
\end{equation}

after simplifications, 

\begin{equation}
	E_{b} = \dfrac{E_{tot}^{X_{12}} - 12 E_{tot}^{X-free} + E_{tot}^{X_{12}TM} - E_{tot}^{X_{12}-frozen} - E_{tot}^{TM-free} + E_{tot}^{X_{12}-frozen} - E_{tot}^{X_{12}}}{13},
\end{equation}

which results in equation,

\begin{equation}
	E_{b} = \frac{E_{tot}^{X_{12}TM} - 12 E_{tot}^{X-free} - E_{tot}^{TM-free}}{13},
\end{equation}

Therefore, both equations are equivalent and must produce the same results.

\newpage

\section{Properties of TM Bulks}

To limit the scope, we also calculate bulk properties of the Transition Metals. The Properties such as cohesive energy ($E_{\text{coh}}$) and atomic radius ($R_{\text{TM}}$) for the lowest-energy unit cell are shown in the graphic below, compared with experimental data.

\begin{figure}[H]
\centering
\includegraphics[scale=0.75]{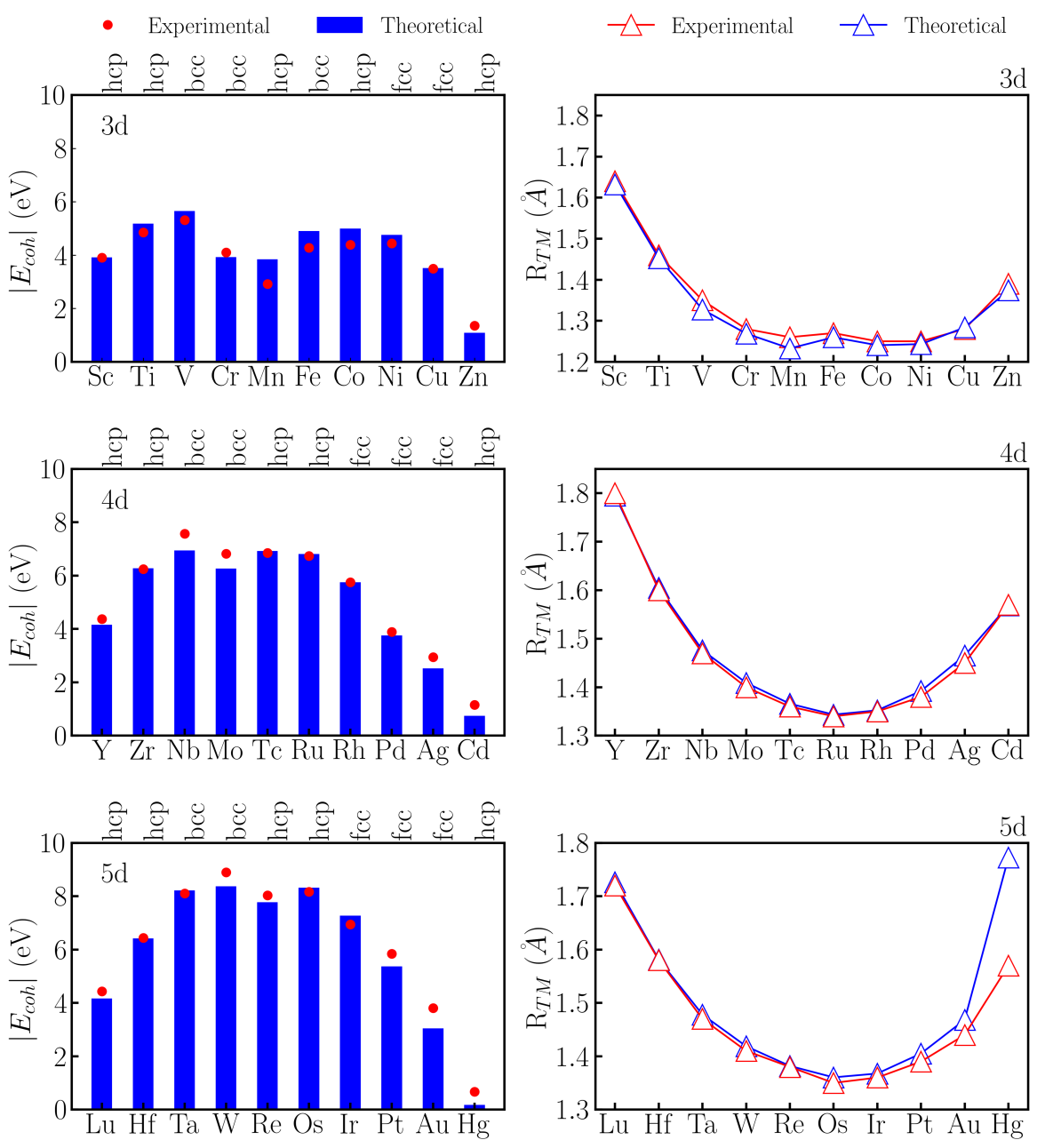}
\caption*{Figure S1: The bulks properties, $E_{\text{coh}}$ and $R_{\text{TM}}$, for TMs as function of the atomic number. In color blue, the theoretical data, and in color red, the experimental data.}
\label{bulks}
\end{figure}

\newpage

\section{Properties of Nanoclusters X$_{12}$TM}

\subsection{Complete Set of Optimized Nanoclusters}

Below, the optimized 13-atom nanoclusters are shown in Figures:

\begin{figure}[H]
\centering
\includegraphics[scale=0.08]{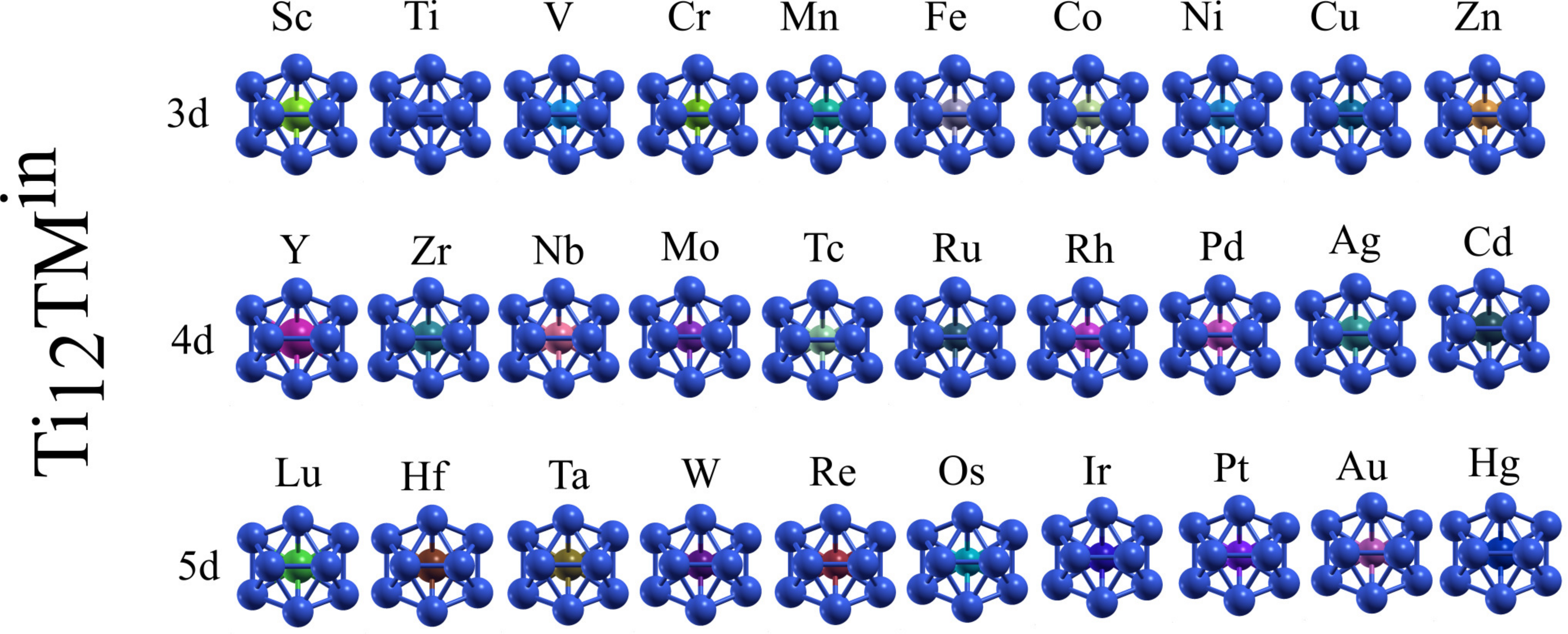}
\caption*{Figure S2: Optimized 13-atom nanoclusters Ti$_{12}$TM$^{in}$ in geometric structure icosahedral (ICO) as in Periodic Table.}
\end{figure}

\begin{figure}[H]
\centering
\includegraphics[scale=0.075]{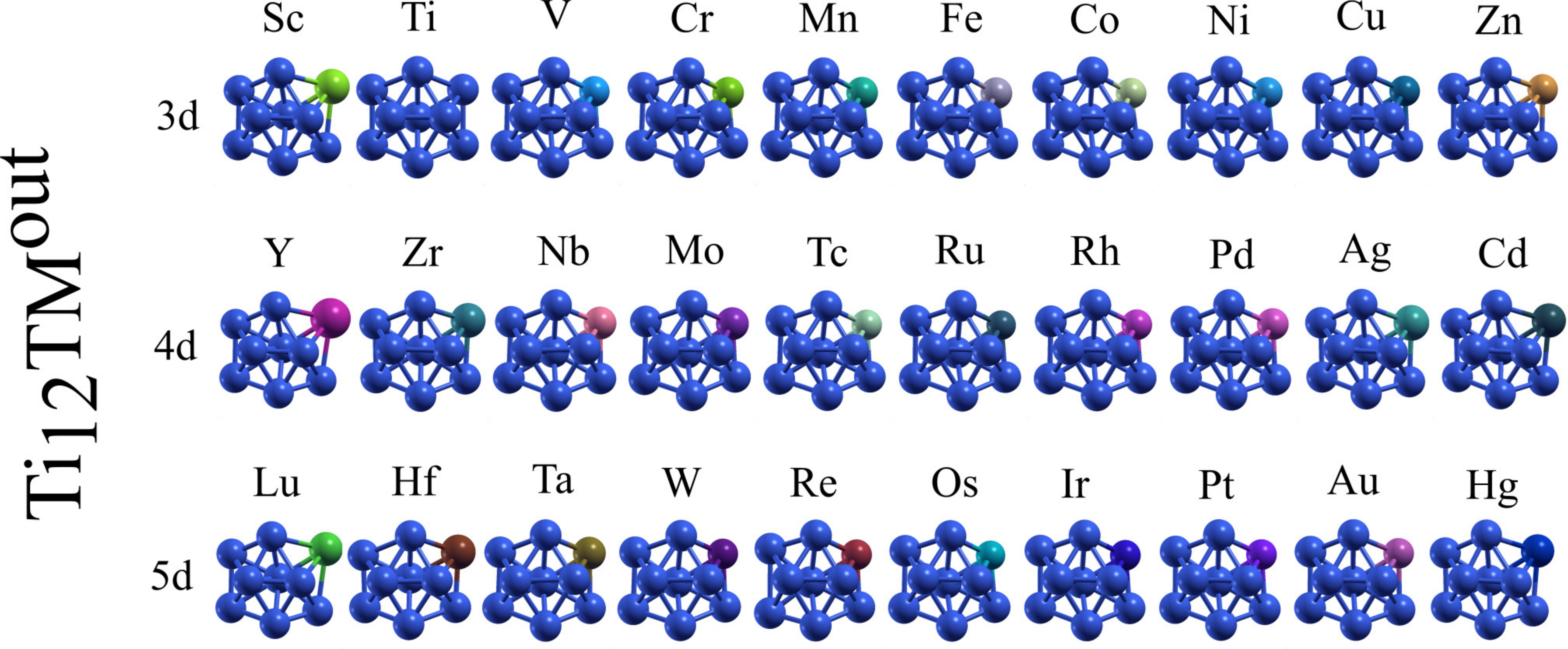}
\caption*{Figure S3: Optimized 13-atom nanoclusters Ti$_{12}$TM$^{out}$ in geometric structure icosahedral (ICO) as in Periodic Table.}
\end{figure}

\newpage

\begin{figure}[H]
\centering
\includegraphics[scale=0.08]{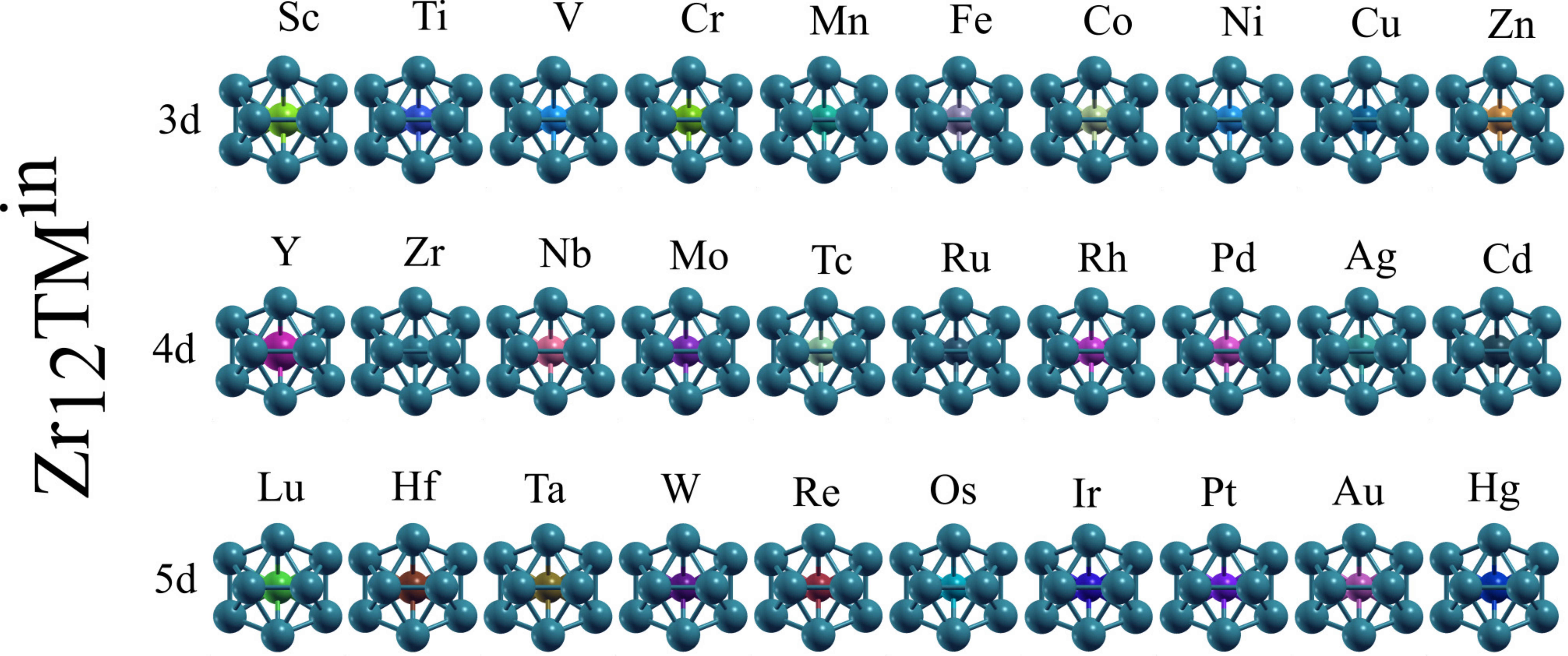}
\caption*{Figure S4: Optimized 13-atom nanoclusters Zr$_{12}$TM$^{in}$ in geometric structure icosahedral (ICO) as in Periodic Table.}
\end{figure}

\begin{figure}[H]
\centering
\includegraphics[scale=0.08]{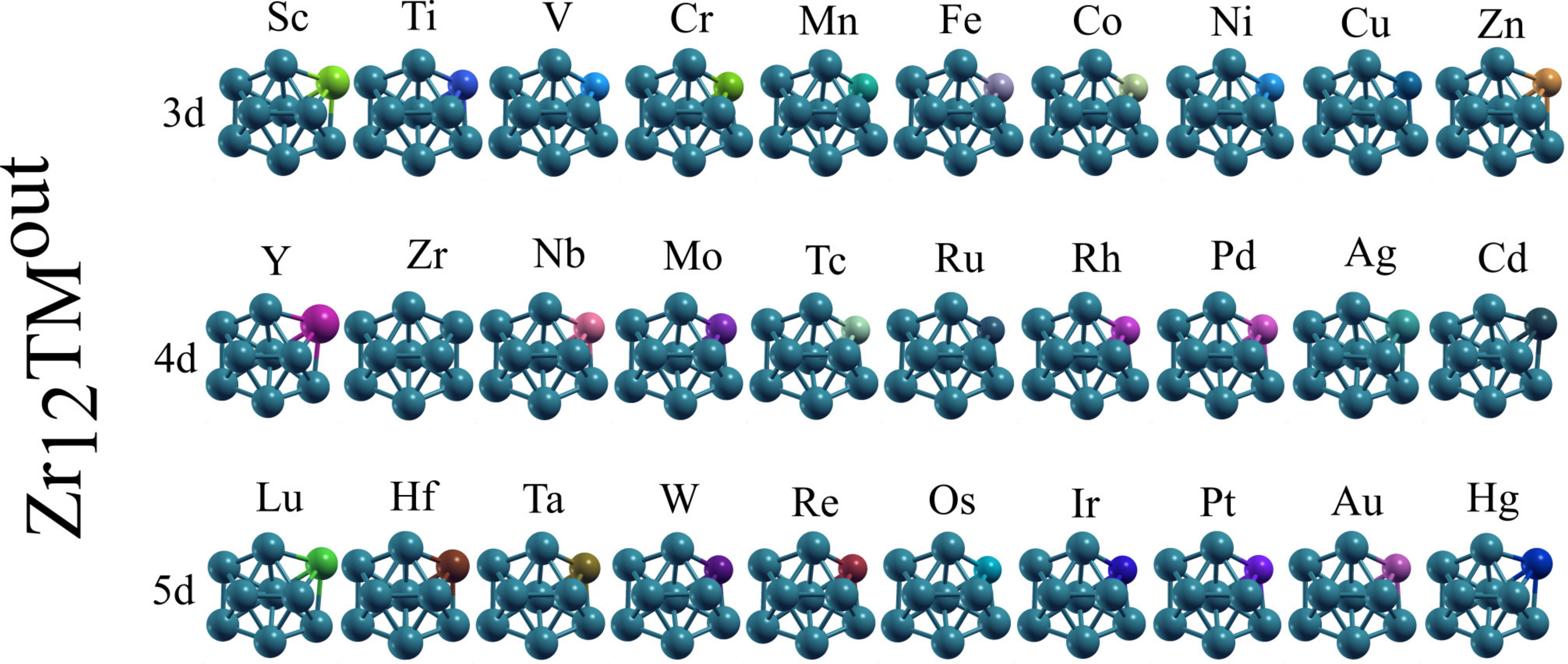}
\caption*{Figure S5: Optimized 13-atom nanoclusters Zr$_{12}$TM$^{out}$ in geometric structure icosahedral (ICO) as in Periodic Table.}
\end{figure}

\newpage

\begin{figure}[H]
\centering
\includegraphics[scale=0.08]{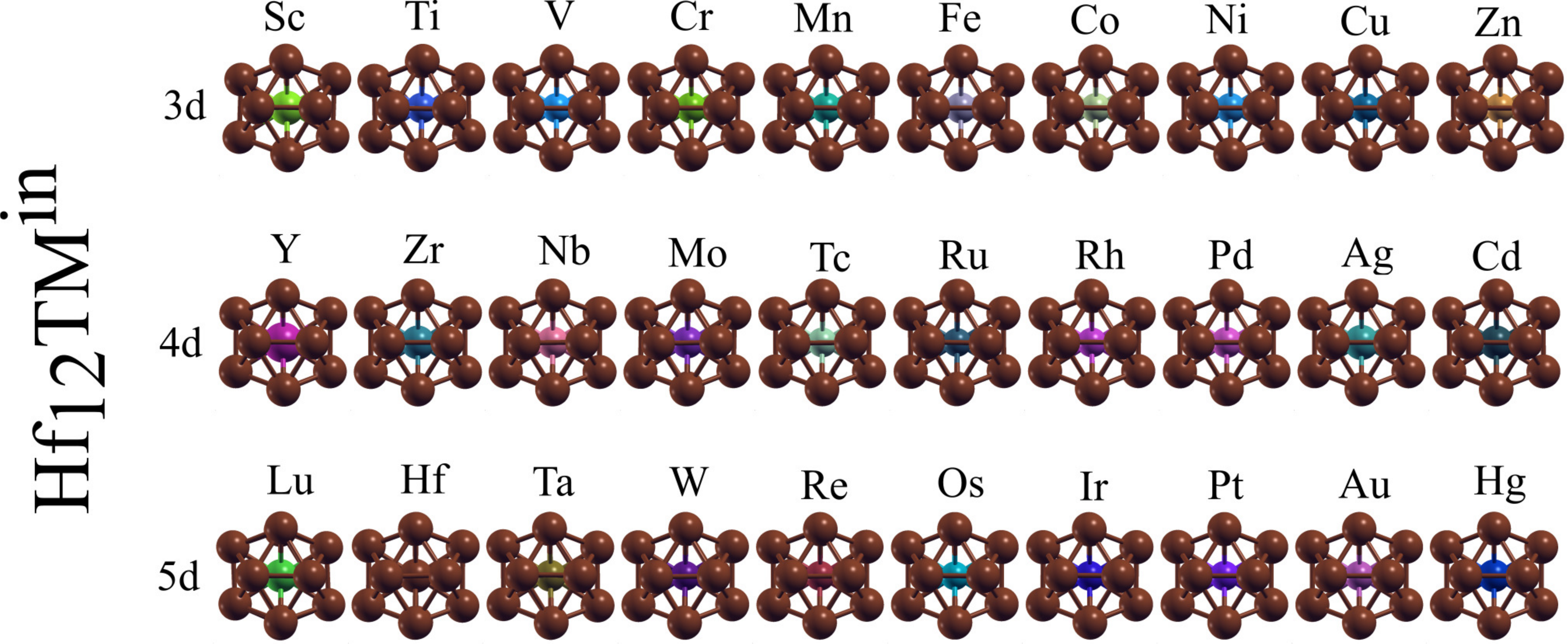}
\caption*{Figure S6: Optimized 13-atom nanoclusters Hf$_{12}$TM$^{in}$ in geometric structure icosahedral (ICO) as in Periodic Table.}
\end{figure}

\begin{figure}[H]
\centering
\includegraphics[scale=0.08]{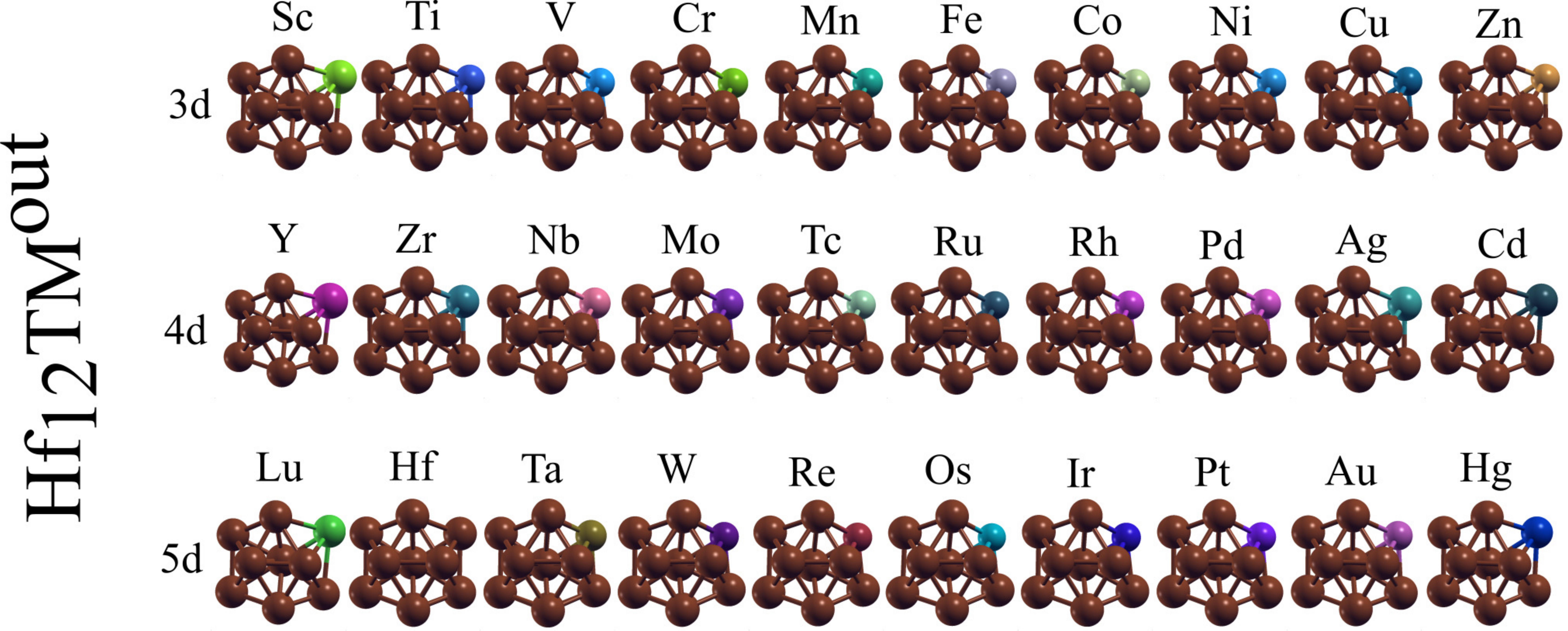}
\caption*{Figure S7: Optimized 13-atom nanoclusters Hf$_{12}$TM$^{out}$ in geometric structure icosahedral (ICO) as in Periodic Table.}
\end{figure}

\newpage

\begin{figure}[H]
\centering
\includegraphics[scale=0.08]{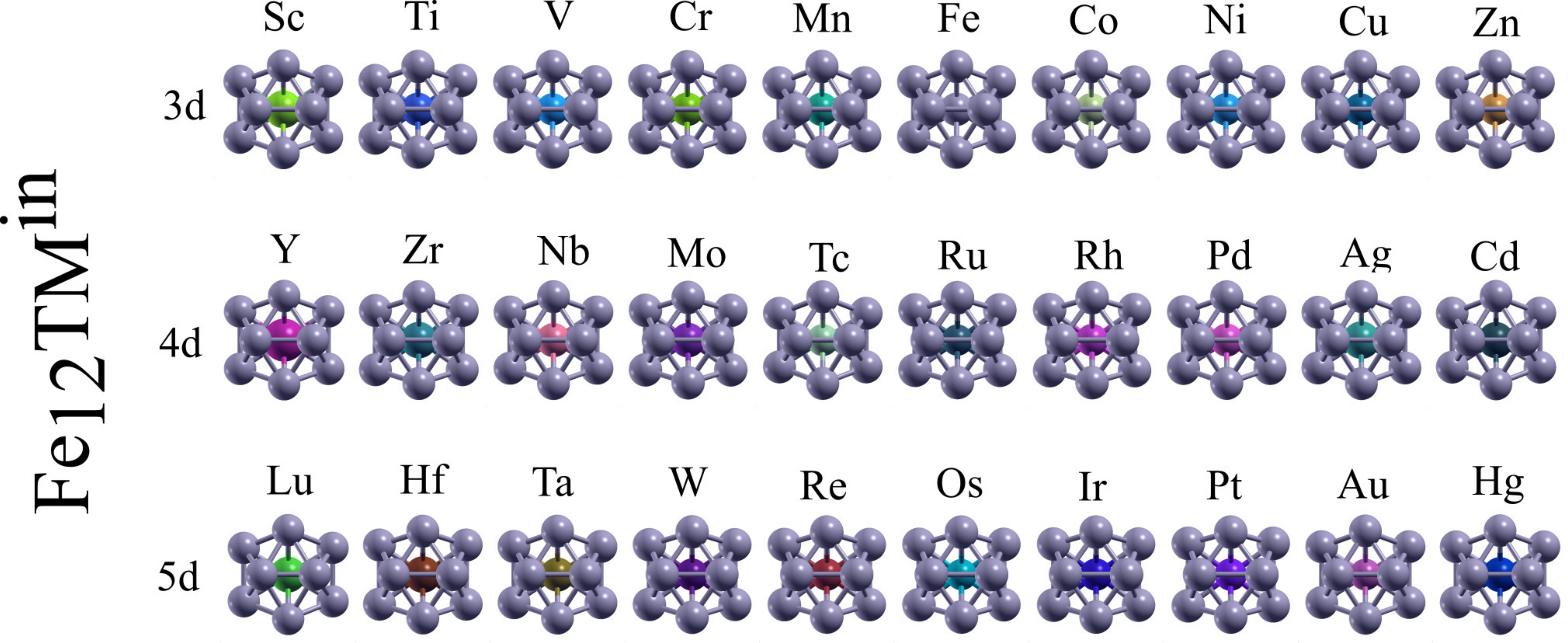}
\caption*{Figure S8: Optimized 13-atom nanoclusters Fe$_{12}$TM$^{in}$ in geometric structure icosahedral (ICO) as in Periodic Table.}
\end{figure}

\begin{figure}[H]
\centering
\includegraphics[scale=0.08]{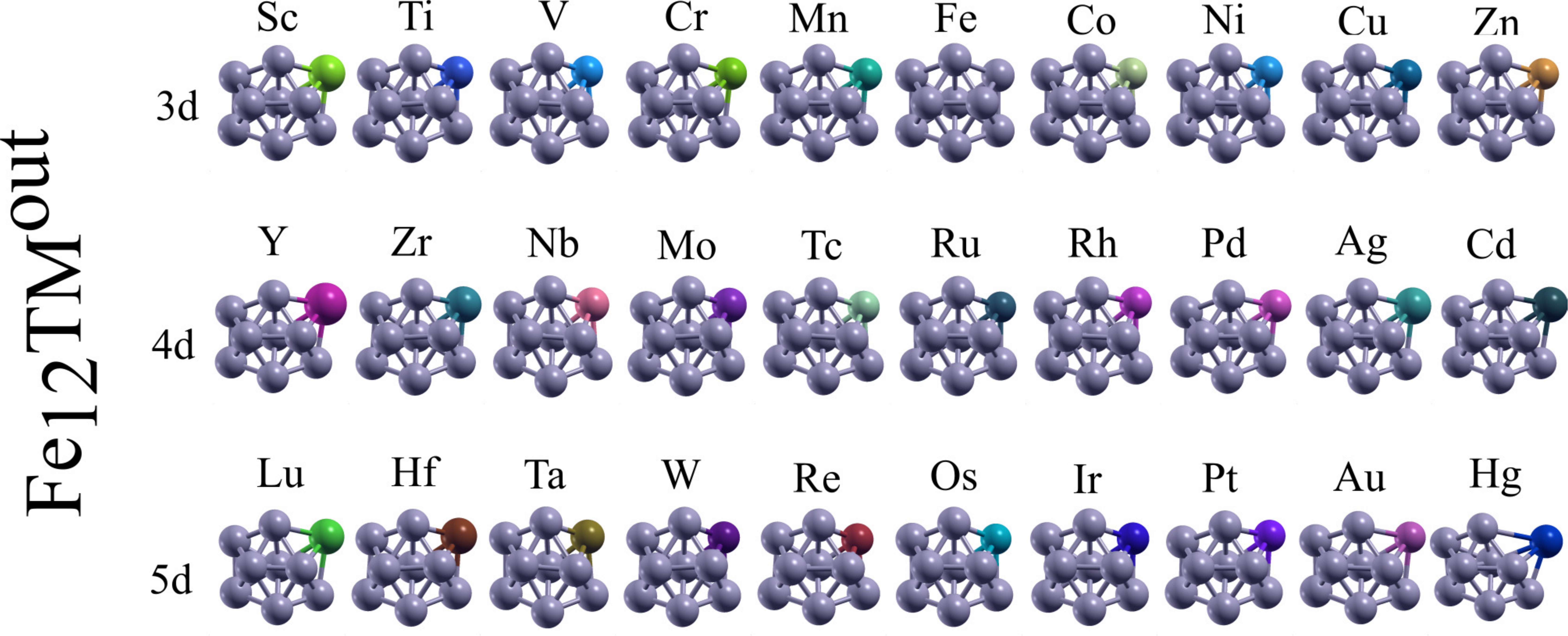}
\caption*{Figure S9: Optimized 13-atom nanoclusters Fe$_{12}$TM$^{out}$ in geometric structure icosahedral (ICO) as in Periodic Table.}
\end{figure}

\newpage

\subsection{Effective Coordination Number and Average Bond Length}

The effective coordination number (ECN) and average bond length (d$_{av}$) were employed to analyze the X$_{12}$TM ICO-like nanoclusters.\cite{Hoppe_25_1970, Hoppe_23_1979} The usual coordination number (CN) defines the number of neighboring atoms to the atom $j$ and attributes the same weight for all atoms $i$ closer to the coordination sphere. This method does not account for bonds that are longer than the cutting radius; therefore, it does not accurately represent distorted structures. In the ECN, the cutting radius is not necessary, since the weight is attributed to every atom $i$ according to the bond length. In this sense, the atom $j$ is surrounded by atoms $i$ with different distances and different weights. Consequently, the weight depends directly on the d$_{av}$, with values smaller (larger) than d$_{av}$ contributing with a weight larger (smaller). Thus, ECN can represent distorted structures while still yielding the correct value for bulk systems. The ECN$_i$ is obtained by
\begin{equation}
    \text{ECN}_i = \sum_j exp\left[ 1 - \left(\dfrac{d_{ij}}{d^i_{av}} \right)^6 \right],
\end{equation}
\noindent where $d_{ij}$ is the atomic distance between $i$ and $j$, $d_{av}$ is defined as
\begin{equation}
    d_{av}^i = \dfrac{\sum_j d_{ij} exp\left[ 1 - \left(\dfrac{d_{ij}}{d_{av}^i} \right)^6 \right] }{ \sum_j exp\left[ 1 - \left(\dfrac{d_{ij}}{d_{av}^i} \right)^6 \right] }.
\end{equation}
The $d_{av}^i$ is obtained self-consistently, where $|d_{av}^i(new) - d_{av}^i(old)| < 0.0001$ is the convergence parameter. The initial value is defined as the smallest bond length between the atom $i$ and all $j$ atoms ($d_{min}^i$). The final value of $d_{av}^i$ is obtained after a few interactions. The average ECN and $d_{av}$ values are obtained by 
\begin{equation}
    \text{ECN} = \dfrac{1}{N} \sum_{i=1}^N \text{ECN}_i,
\end{equation}
\begin{equation}
    d_{av} = \dfrac{1}{N} \sum_{i=1}^N d_{av}^i~,
\end{equation}
\noindent where $N$ is the total number of atoms in the nanocluster.

\newpage

\subsection{Vibrational Frequencies}

Vibrational frequencies were calculated for systems Ti, Zr, and Hf. This result shows that systems are local minima with honest and positive frequencies.

\begin{figure}[H]
\centering
\includegraphics[scale=0.4]{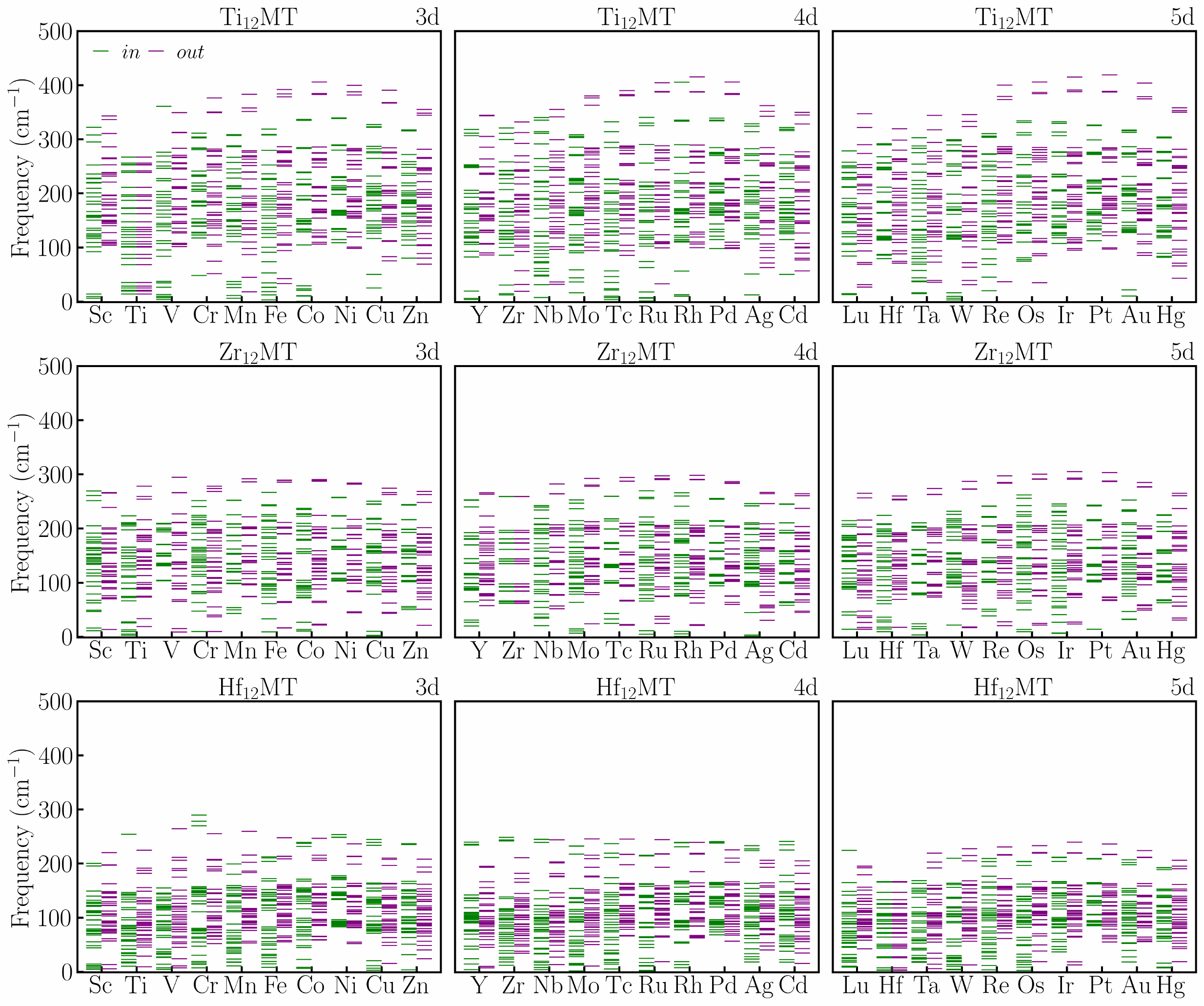}
\caption*{Figure S10: Vibrational frequencies for nanoclusters of Ti, Zr, and Hf as a function of the atomic number. In purple are systems \textit{in} and green are systems \textit{out}.}
\label{ebzh}
\end{figure}

\newpage




\subsection{Formation Energy}

\begin{figure}[H]
\centering
\includegraphics[scale=0.4]{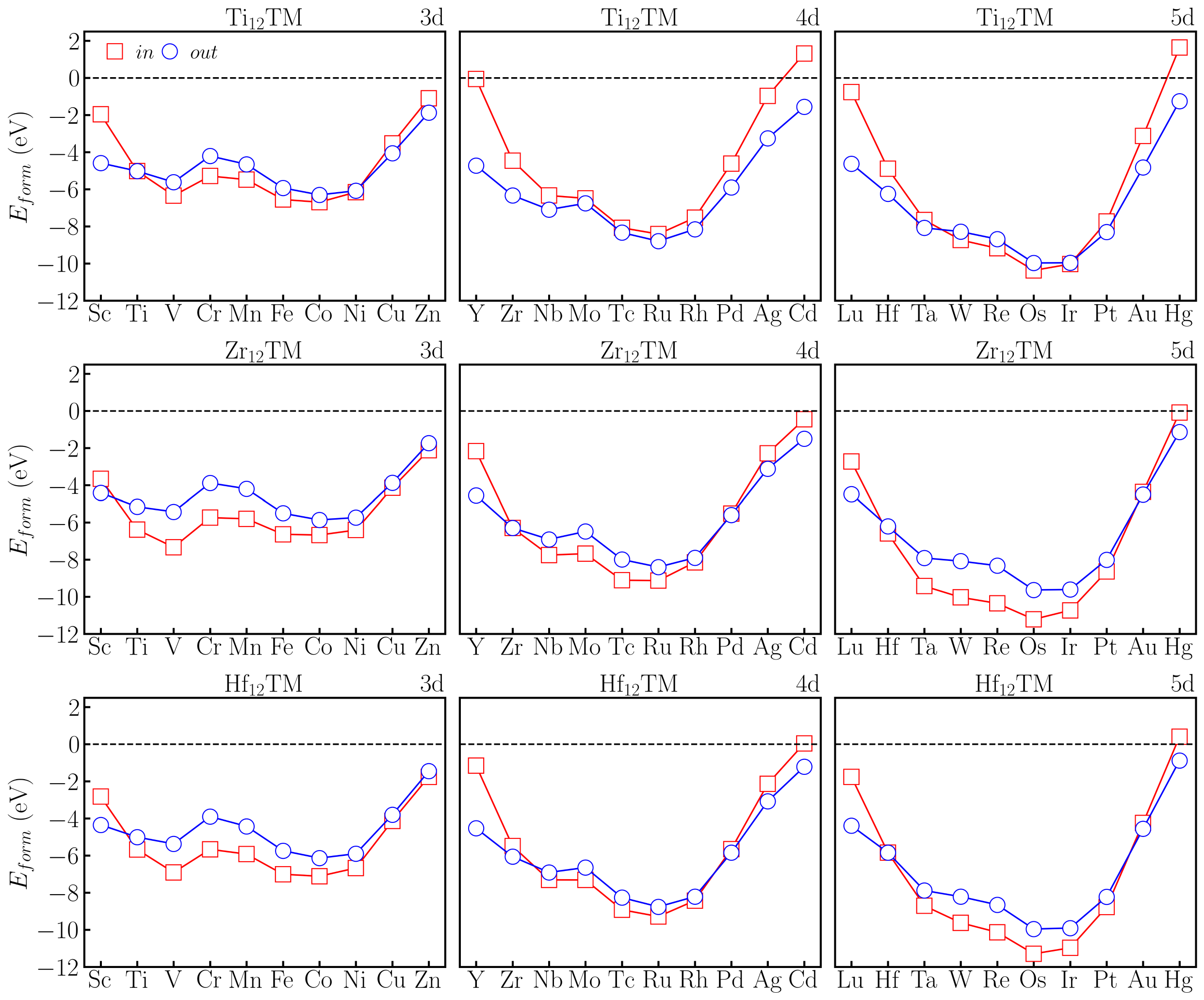}
\caption*{Figure S11: Formation Energy for nanoclusters of Ti, Zr, and Hf as a function of the atomic number. In color, blues are the \textit{in} systems, and blue is the \textit{out} system.}
\label{ebzh}
\end{figure}

\newpage

\subsection{Binding and Formation Energies for Fe-based Systems}

\begin{figure}[H]
\centering
\includegraphics[scale=0.4]{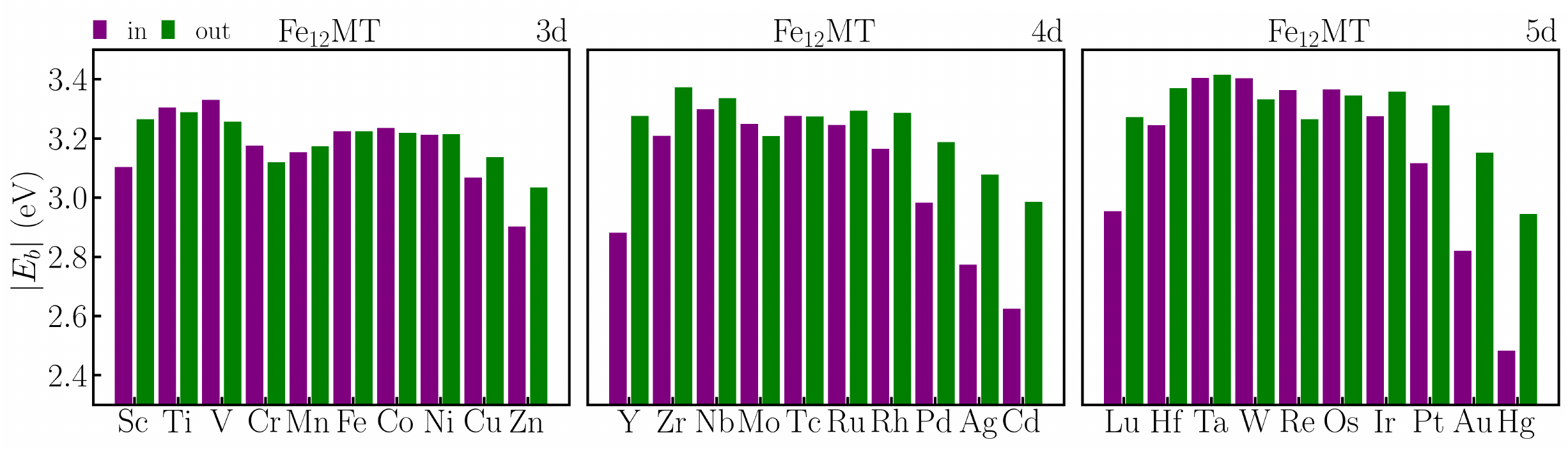}
\caption*{Figure S12: Binding energy for nanoclusters of Fe as a function of the atomic number. In color purple are the systems \textit{in} and green the \textit{out} systems.}
\label{ebfew}
\end{figure}

\begin{figure}[H]
\centering
\includegraphics[scale=0.4]{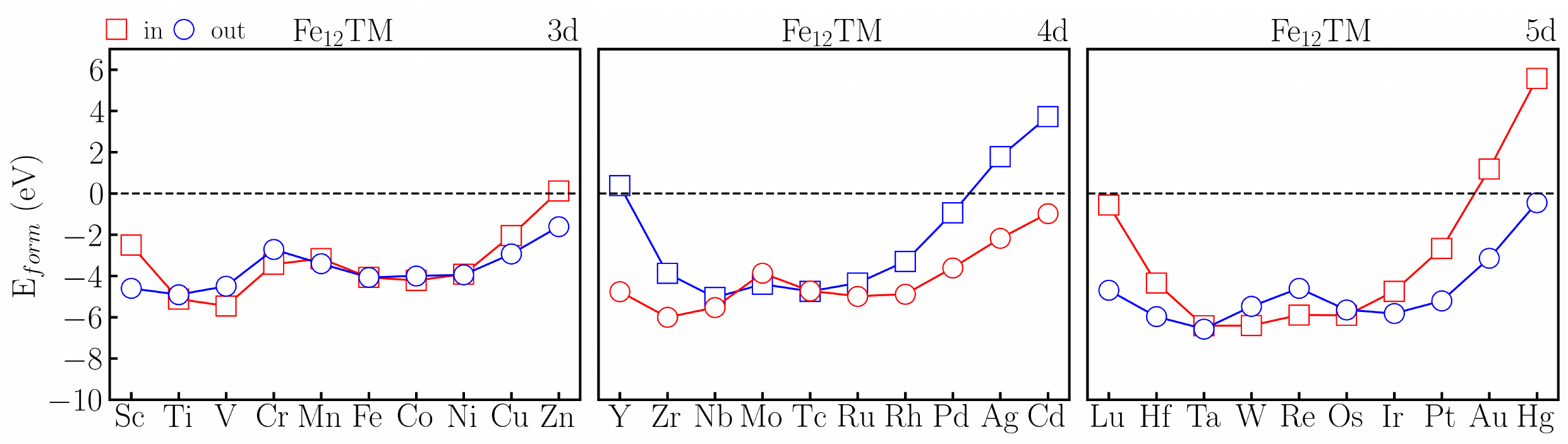}
\caption*{Figure S13: Formation Energy for nanoclusters of Fe as a function of the atomic number. In color, blues are the \textit{in} systems, and blue is the \textit{out} system.}
\label{effe}
\end{figure}

\newpage

\subsection{Geometric Properties for Fe-based Systems}

\begin{figure}[H]
\centering
\includegraphics[width=\linewidth]{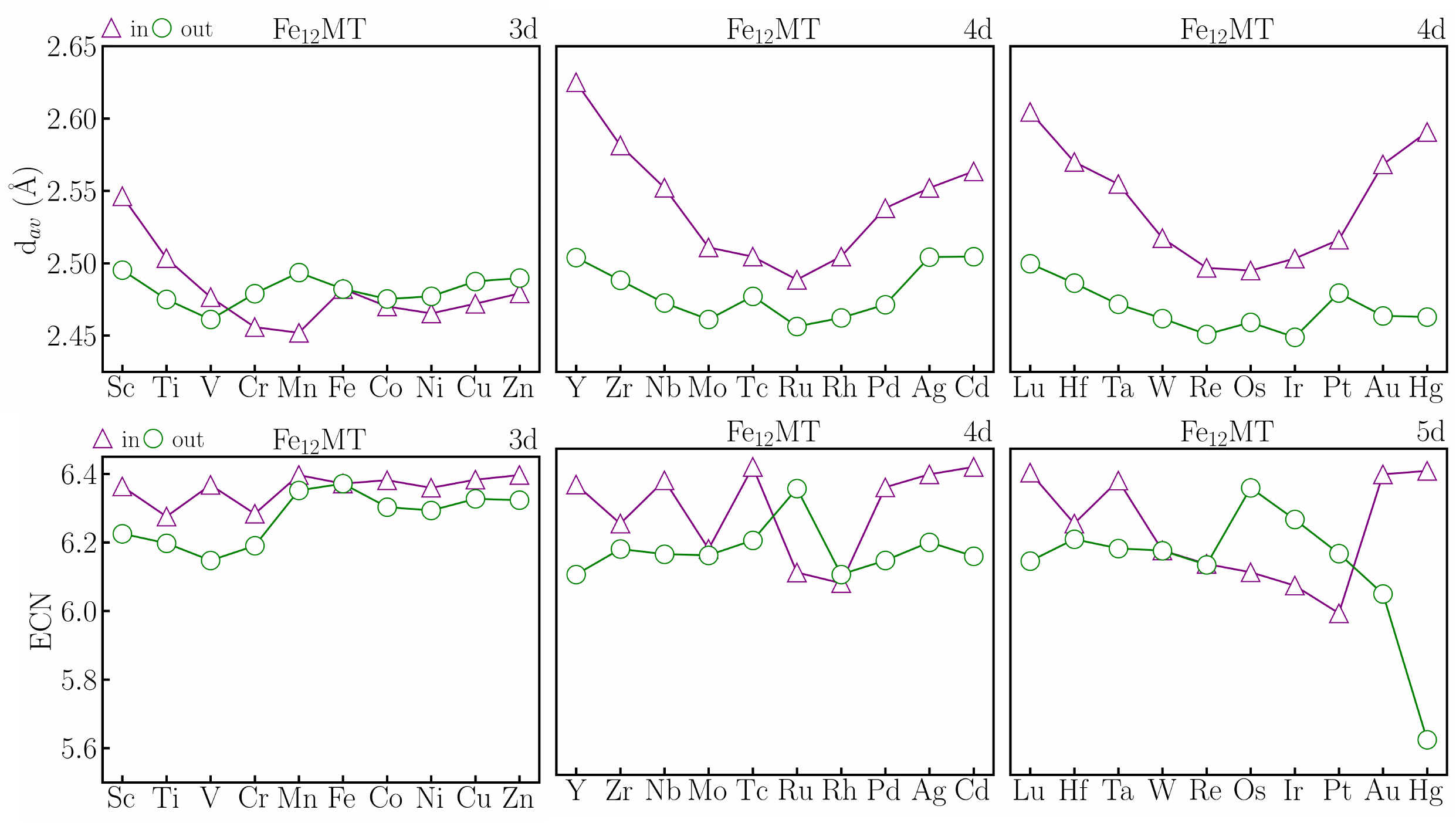}
\caption*{Figure S14: d$_{av}$ and ECN for nanoclusters of Fe. In color purple are the in systems, and in green are the out systems.}
\label{davecnfe}
\end{figure}

\newpage

\subsection{Center of Gravity of the Occupied $d$-states}

\begin{figure}[H]
\centering
\includegraphics[scale=0.4]{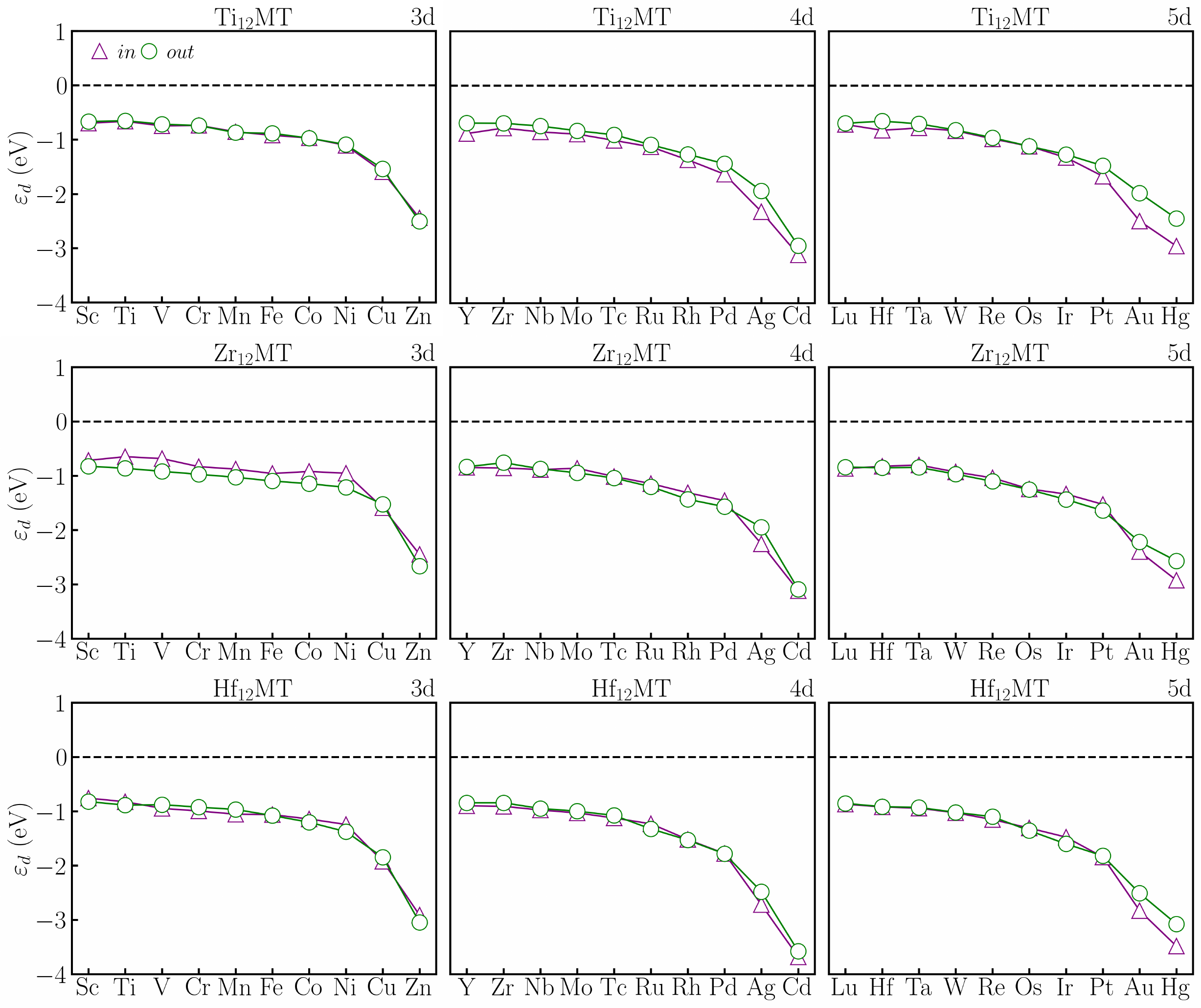}
\caption*{Figure S15: Center of gravity of the occupied $d$ states for nanoclusters of Ti, Zr, and Hf as a function of the atomic number. In purple are the \textit{in} systems, and in green the \textit{out} systems.}
\label{ebzh}
\end{figure}

\newpage

\subsection{HOMO-LUMO Gap}

\begin{figure}[H]
\centering
\includegraphics[scale=0.4]{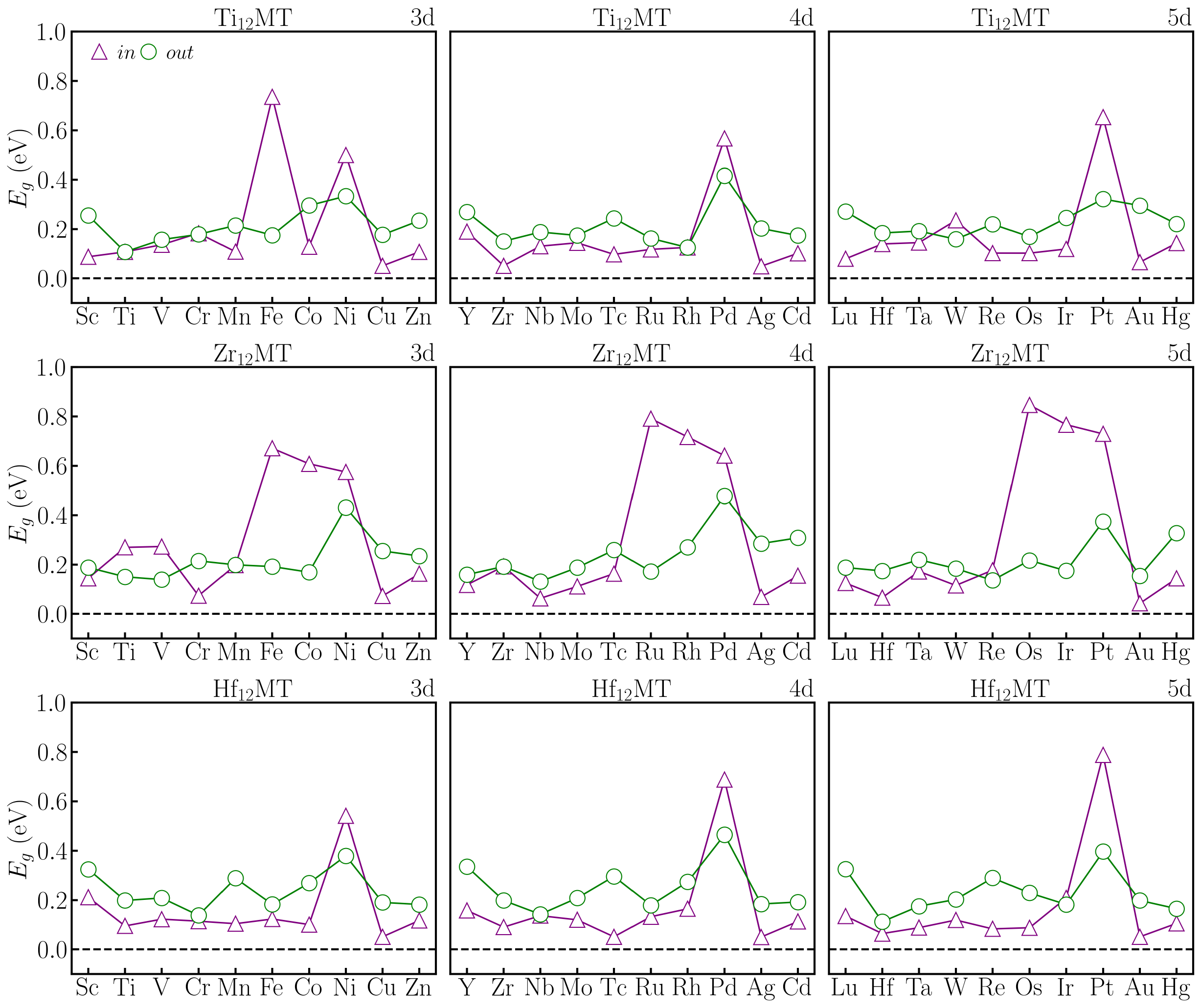}
\caption*{Figure S16: HOMO-LUMO gap for nanoclusters of Ti, Zr, and Hf as a function of the atomic number. In purple are the \textit{in} systems, and in green the \textit{out} systems.}
\label{ebzh}
\end{figure}

\newpage

\begin{figure}[H]
\centering
\includegraphics[scale=0.4]{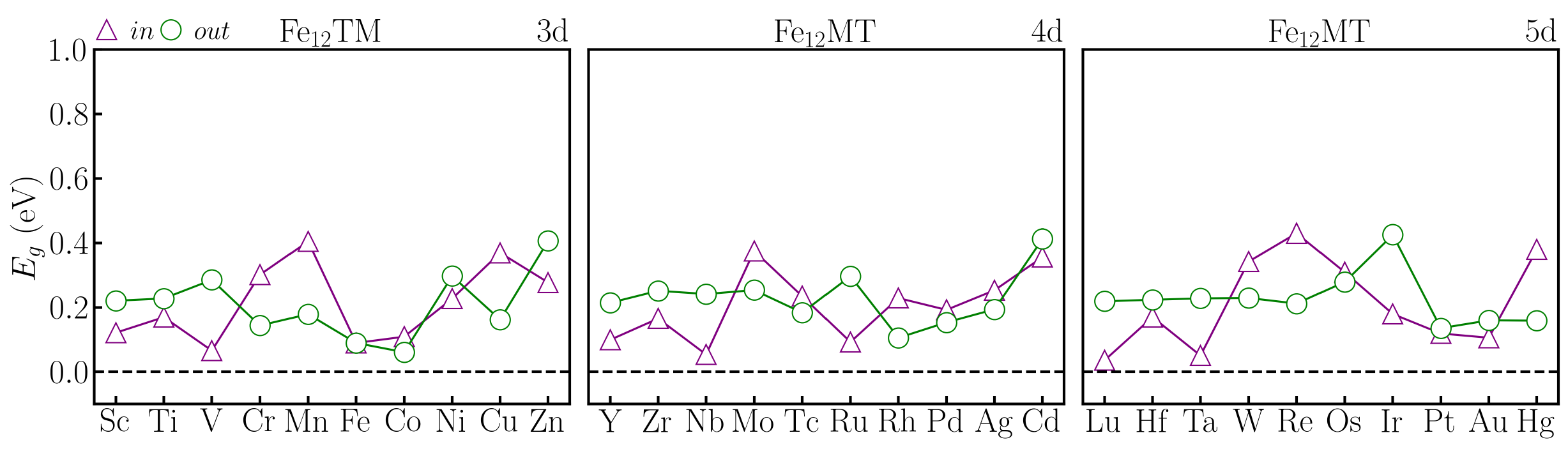}
\caption*{Figure S17: HOMO-LUMO gap for nanoclusters of Fe as a function of the atomic number. In purple are the \textit{in} systems, and in green the \textit{out} systems.}
\label{hlfe}
\end{figure}

\newpage

\subsection{Validation for Fe-based Systems}

To critically validate the generality of our trends and the predictive capability of our ML model, we selected \ce{Fe$_{12}$TM} nanoclusters as a test case and analyzed both $in$ and $out$ configurations. The choice of \ce{Fe} as a host system is particularly justified given the robust literature indicating that \ce{Fe$_{13}$} adopts an ICO-like geometry as its ground state, consistent with earlier studies.\cite{Piotrowski_155446_2010,Chaves_15484_2017} 

Below, we present a direct comparison between DFT-calculated and ML-predicted $E_{form}$ for \ce{Fe$_{12}$TM} nanoclusters across the \num{3}$d$, \num{4}$d$, and \num{5}$d$ TM series. Both $in$ and $out$ configurations are depicted, with excellent agreement between ML and DFT predictions, even for dopants in the unseen (validation) set, highlighting again the model's predictive capacity and transferability. The accurate reproduction of trends of stability, i.e., the increased stabilization of $in$ configurations towards mid-series TMs (e.g., \ce{Mn}, \ce{Fe}, \ce{Co}) and the trend to $out$ configurations at the series boundaries, highlights the model's capability to capture small electronic and structural effects.

\begin{figure}[H]
\centering
\includegraphics[width=0.95\linewidth]{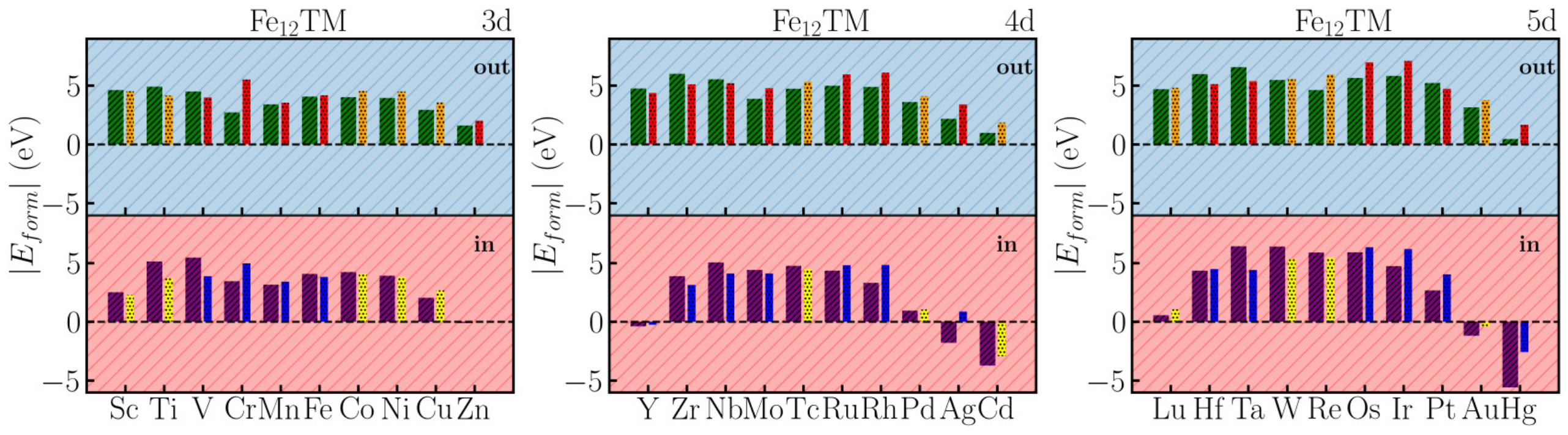}
\caption*{Figure S18: Comparison of DFT (solid bars) and (seen and unseen) ML-predicted (hatched bars) formation energies ($|E_{form}|$) for \ce{Fe$_{12}$TM} nanoclusters with \num{3}$d$, \num{4}$d$, and \num{5}$d$ TM dopants. Top and bottom panels show $out$ and $in$ configurations, respectively. The high consistency across series and configurations, including unseen validation cases, corroborates the reliability and transferability of the ML model.}
\end{figure}

Consequently, our results for \ce{Fe$_{12}$TM} nanoclusters also strongly confirm the generalizability of the group IV-based nanocluster structural and energetic design principles developed. In addition, the even better agreement between ML predictions and DFT calculations in Figure S18 further confirms the predictive capability and universal applicability of our model across hosts with varying atomic sizes, electronic structures, and magnetic properties.

\newpage

\section{Predictive Artificial Intelligence}

\begin{figure*}[!ht]
\centering
\includegraphics[width=0.95\linewidth]{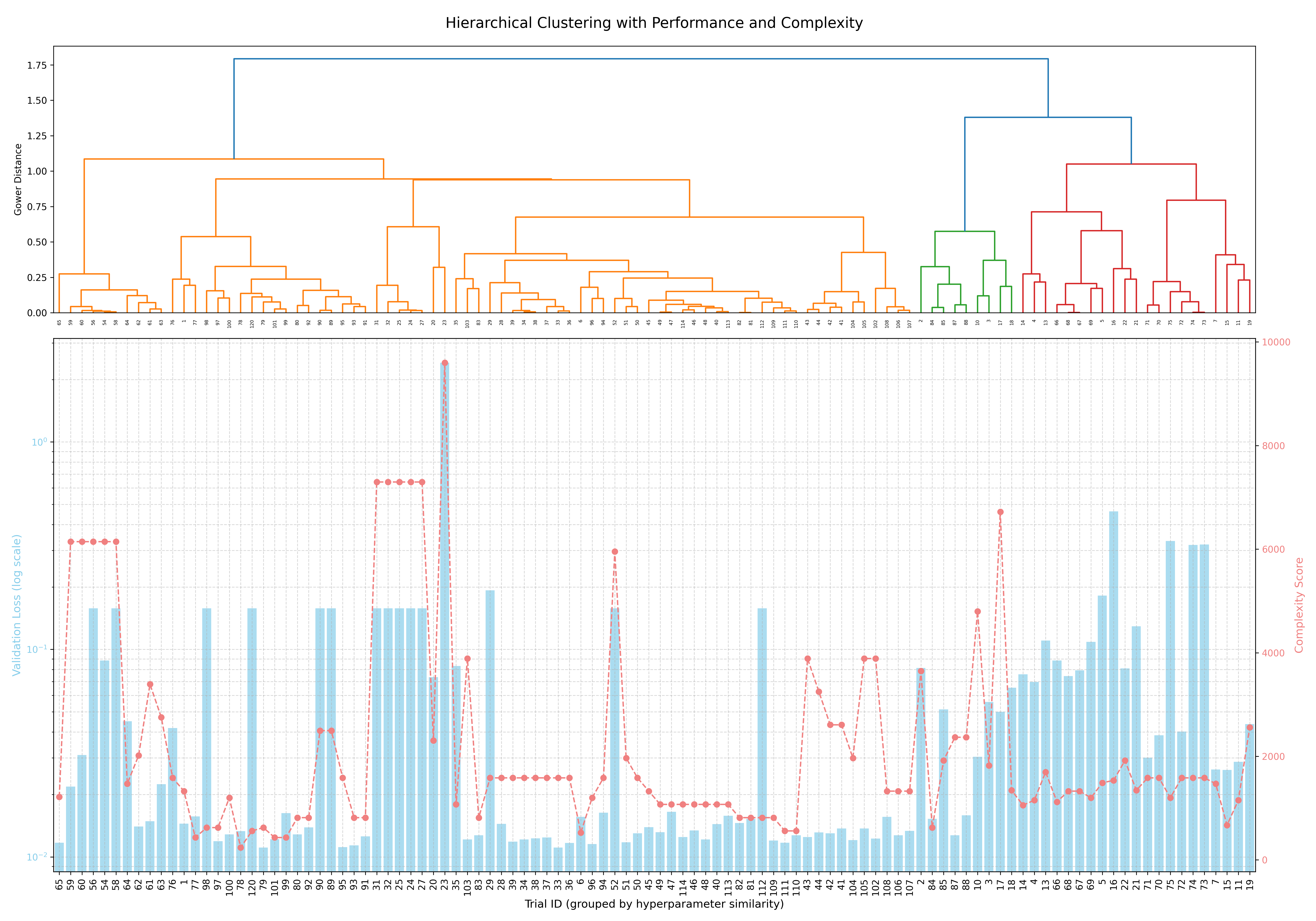}
\caption*{Figure S19: Hierarchical clustering of hyperparameter trials and their performance–complexity trade-off. The upper panel shows a dendrogram built from the Gower distance between all FTTransformer hyperparameter configurations explored during the search, grouping trials with similar architectural and training settings. The lower panel reports, for the exact trial ordering, the validation loss (blue bars, left $y$-axis, logarithmic scale) and the corresponding model–complexity score (red dashed line with markers, right $y$-axis), highlighting clusters of configurations that simultaneously minimize loss while avoiding unnecessarily complex models.}
\label{fig:hierarchical_performance_complexity}
\end{figure*}

Figure S19 summarizes the structure of the hyperparameter search space and its relationship to model performance and complexity. Hierarchical clustering based on the Gower distance organizes all trials into families of similar architectures and optimization settings, which appear as contiguous blocks along the $x$-axis. Plotting the validation loss and complexity score for this clustered ordering reveals clear Pareto-like regions: some clusters contain overly simple models with high loss, others contain highly complex models with only marginal performance gains, and a narrow band of intermediate-complexity models achieves the best trade-off. The final FTTransformer configuration used in this work was selected from this latter region, combining low validation loss with moderate complexity.

\bibliography{jshort.bib,boxref.bib}